\documentclass[journal=bichaw,manuscript=article]{achemso}

\usepackage{chemformula} 
\usepackage[T1]{fontenc} 

\usepackage{upgreek}
\usepackage{graphicx} 
\usepackage{subfigure}
\usepackage{epstopdf}

\author{Ly H. Nguyen}
\affiliation[KL]{Key Laboratory for Multiscale Simulations of Complex Systems, VNU University of Science,
Vietnam National University, Hanoi,\\
 334 Nguyen Trai street, Thanh Xuan district, Hanoi, Vietnam}
\author{Tuyen T. Tran}
\affiliation[KL]{Key Laboratory for Multiscale Simulations of Complex Systems, VNU University of Science,
Vietnam National University, Hanoi,\\
 334 Nguyen Trai street, Thanh Xuan district, Hanoi, Vietnam}
 \author{Truong Thi Ngoc Lien}
 \affiliation[HUST]{Hanoi University of Science and Technology, \\
 1 Dai Co Viet street, Bach Khoa, Hai Ba Trung district, Hanoi, Vietnam}
 \author{Mai Hong Hanh}
\affiliation[KL]{Key Laboratory for Multiscale Simulations of Complex Systems, VNU University of Science,
Vietnam National University, Hanoi,\\
 334 Nguyen Trai street, Thanh Xuan district, Hanoi, Vietnam}
\author{Toan T. Nguyen}
\email{toannt@vnu.edu.vn, toannt@hus.edu.vn}
\affiliation[KL]{Key Laboratory for Multiscale Simulations of Complex Systems, VNU University of Science,
Vietnam National University, Hanoi,\\
 334 Nguyen Trai street, Thanh Xuan district, Hanoi, Vietnam}

\title[Zinc$-$finger overcharging]{Overcharging of zinc ion in the structure of zinc$-$finger protein 
is needed for DNA binding stability}


\begin{document}

\begin{abstract}
Zinc finger structure, where a Zn$^{2+}$ ion binds to four 4 cysteines or histidines 
in a tetrahedral structure, is a very common motif of nucleic acid$-$binding proteins.  
The corresponding interaction model is present in 3\% of the genes in human genome. 
As a result, zinc$-$finger has been extremely useful in various therapeutic 
and research capacities, and in biotechnology.  In stable configuration of zinc$-$finger, 
the cysteine amino acids are deprotonated and become
negatively charged. Thus, the Zn$^{2+}$ ion is overscreened by 
four cysteine charges ({\em overcharged}). 
Whether this overcharged configuration is also stable
when such a negatively charged zinc$-$finger binds to a negatively charged DNA molecule
is unknown. We investigated how the deprotonated state of cysteine influences its structure, dynamics, 
and function in binding to DNA molecules by
using an all$-$atom molecular dynamics simulation up to microsecond range of an 
androgen receptor protein dimer. Our results showed that 
the deprotonated state of cysteine residues is essential for mechanical stabilization of 
the functional, folded conformation. Not only this state stabilizes the protein structure, 
it also stabilizes the protein$-$DNA binding complex. The differences in structural and energetic 
properties of the two sequence-identical monomers are also investigated 
and show the strong influence of DNA on the structure of zinc-fingers protein dimer upon complexation.
Our result can potentially lead to better molecular understanding of one of the most common classes
of zinc fingers.
\end{abstract}

\section{Introduction}

Zinc finger proteins are among the most abundant proteins in eukaryotic genomes.
These proteins are encoded by 3\% of the human genome \cite{Lander2001, Klug2010, Kluska2018}.
Their functions are extraordinarily diverse and include DNA recognition, 
RNA packaging, transcription activation, regulation of apoptosis, 
protein folding and assembly, and lipid binding. 
%For example, 
There are increasing evidence
the potential roles of zinc finger in cancer progression \cite{jen2016}.  The aberrant expression of 
C2H2 zinc finger proteins contributes to tumorigenesis in many different aspects.
Another example is their chaperon function in the nucleocapsid protein
of the human immunodeficiency virus type 1 (HIV-1) \cite{Mitra2013}. 
This protein plays an important role in the life cycle
of this virus and has been an attractive target for therapeutic treatment.
In biotechnology, their sequence$-$specific DNA$-$binding property is also used 
in bio$-$engineering to target desired DNA genome sequences \cite{Jamieson2003}.
For example, the prostate$-$specific antigen (PSA) protein, which has zinc$-$fingers 
for nucleic acid binding, 
is a common marker for prostate cancer \cite{balk2003biology}. 
Therefore, one can detect the presence of PSA in a sample by using 
a substrate that is functionalized with aptamers (short DNA molecules) that only the 
PSA protein can recognize specifically 
\cite{formisano2015optimisation,aus2005individualized,botchorishvili2009,lilja2008,liu2012detection,lai2006}. 
The electrochemical properties of the substrate will change upon binding of PSA proteins to the aptamers
and can be measured accurately using a companion electric circuit. The strength of the perturbation 
is a measure of the PSA concentration in the sample. Thus, 
PSA concentration can be detected and measured rather
 accurately, allowing for the early detection of prostate cancer.

\begin{figure*}
	\centering
	\includegraphics[width=14cm]{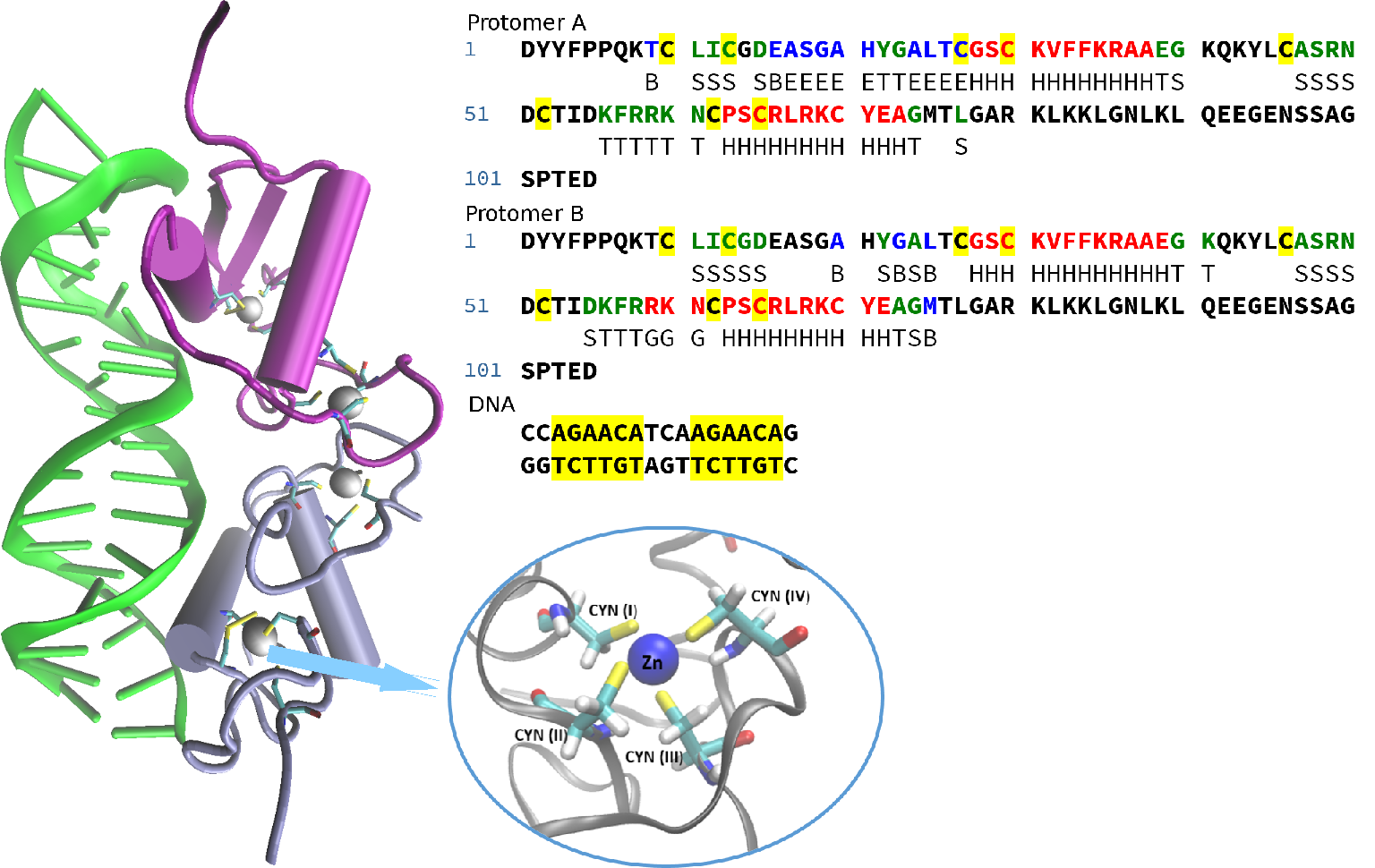}
	\caption{Structure of the zinc$-$finger complex studied in this work: 
DNA (green), protein dimer (purple - protomer A, red - protomer B)
and four Zinc ions (silver ball).  Zoomed to one of the zinc ions, four cysteine residues 
coordinate with the zinc ion in a tetrahedral structure. Top right is the amino
acid sequence of individual protomers with their secondary structure (DSSP classification) listed below. 
The cysteine residues that make up the four zinc$-$finger are highlighted in yellow. 
The nucleic acid base sequence of the DNA segment with repeated upstream and downstream patterns
are also highlighted in yellow. } 
	\label{fig:system}
\end{figure*}

Zinc finger structures are as diverse as their functions. However, the most common structure
follows the same motif of a short $\alpha$-helix, two $\beta$-strands and a loop
\cite{Laity2001}. The amino acid residues of this protein segment are arranged 
in three-dimensional space such that the zinc ion would coordinate with 
4 residues, Cys2His2, Cys3His or Cys4, 
to maintain the rigidity of the structure. 
The helix group then binds to the major groove of the DNA double helix. 
The rest of the residues form hydrogen bonds 
to appropriate nucleic acid residues in a sequence specific manner.
This genome specificity makes zinc$-$finger, either natural or artificially engineered, 
a very promising molecule for biotechnological 
application for gene therapy or recognition.
Therefore, understanding the structure and functions of zinc$-$finger proteins,
especially at the molecular level, is very important for biological, biotechnological and bioengineering
applications \cite{Wolfe2000}.

In this work, we focus on investigating the structures, stability and DNA-interaction mechanism
of the androgen receptor DNA$-$binding domain \cite{shaffer2004} (see Fig. \ref{fig:system})
using molecular dynamics simulations. Several recent computational
studies have been conducted on zinc finger proteins \cite{Godwin_2017,Godwin2015, Hamed2016,Lee2010a} 
with different focuses. In the present study, the androgen receptor DNA-binding domain
is investigated not only because this is an important protein for prostate cancer biosensor
application mentioned earlier, but also
for several important reasons from biological and physical points of view:

{\em First}, these ZnCys4 proteins are standard, classical fold
$\beta\beta\alpha$ zinc fingers. Therefore, studying this structure can 
potentially elucidate the structure and dynamics of the most common 
class of zinc fingers. Additionally, the experimentally resolved structure  also contains the direct repeat
DNA response element that this protein binds to. This helps with
truthful orientation of protein$-$DNA complex for computational investigation of their
molecular interaction, which is one of the main goals of this work.

{\em Second}, 
this complex has a dimer of proteins, protomer A and protomer B, with identical
amino acid sequences (Fig. \ref{fig:system}). They also bind to identical 'AGAACA' DNA sequences, 
called "upstream" and "downstream" repeat sequences. 
Yet, despite identical amino acid and nucleic acid sequences,
the two protomers have two different, mirroring secondary structures and binding poses. 
This finding is an {\em interesting deviation} from standard concept in biology that sequence determines
structure \cite{cellBook}. 
The secondary structure information for each residue using DSSP classification is shown below 
their sequences in Fig. \ref{fig:system}.
Several $\beta-$strands are absent in this structure. Out of four zinc fingers present, only
one zinc finger of protomer A shows the $\beta-$strands. All the standard $\beta-$strands
of the other fingers have been downgraded to $\beta-$bridge bonds. 
This is clearly due to the change in the secondary structure upon binding of these proteins to DNA.
Thus, investigating this system allows us to understand the influence of the interaction
with DNA on the zinc finger structure at the molecular level.

{\em Third}, previous studies using molecular dynamics simulations \cite{Godwin_2017,Lee2011},
quantum mechanical calculations \cite{Dudev2002,Peters2010}, 
or indirect experiment combined with simulation \cite{Brandt2009}
have suggested that the cysteine amino acids in their electrostatic binding
with zinc ion are not in their natural neutrally charged state but rather in their negatively
charged deprotonated state. 
This case is a very interesting physics problem of overcharging. Indeed, the charge of the 
zinc ion is $+2e$, whereas the total charge of the four deprotonated cysteine amino acids is $-4e$. 
This means that the cysteine charges {\em overcondense} on the zinc ion. Hence, the net charge of 
the zinc ion is negative (overcharged). This finding is especially interesting based on the fact 
that DNA molecule is also negatively charged in aqueous solution. Thus, 
one has the situation where a negative zinc-finger complex binds to negative DNA molecule.
This fact seems to be counter-intuitive from the
electrostatic perspective.

The aim of this work is to understand the structure, interaction and mechanism
of DNA binding of the dimeric zinc finger protein at the molecular level. 
However, this task would require comprehensive investigation in many different aspects. 
Here, we focus on the electrostatics of the zinc ion and how this electrostatics affects 
the structures and free energy of the complex.
Overcharging in biological system happens when the screening charges are of {\em high valence}
 \cite{Grosberg2002,Nguyen2017,Nguyen2016,Hall2009,Netz2001,Gelbart2000,GronbechJensen1997a}. 
In those cases, their mutual electrostatic interaction dominates over the spatial entropy,
thereby resulting in the positional correlation of their distribution on 
a charged surface. This in turns leads to the overcondensation
of these high valence counterions on the surface and the overscreening of its charge. 
The same physics also leads to the phenomenon of like-charge attraction of these surfaces
in the presence of high valence counterions \cite{Grosberg2002,Naji2004}. We argue that a similar physical 
mechanism is applied here. Cysteine amino acids have a charge of only $-e$ and thus cannot
be considered as high$-$valence screening charge. However, their attachment to the protein
polypeptide backbone severely limits their mobility. As a result, cysteines cannot act
as mobile negative charge in screening zinc ion. Hence, their spatial entropy is eliminated.
This leads to the overcharging the zinc $+2e$ ion in the same way
multivalent counterions overscreen charged surface when 
electrostatic interaction dominates over entropy.

Molecular dynamics are carried out for two systems in a setup similar to a previous study
of isolated zinc finger proteins \cite{Godwin_2017} to show the difference 
between undercharged and overcharged states, and 
to stress the influence of protein DNA interaction. The first system,
called the CYN system, is the overcharged zinc-finger where the cysteine amino acids
were deprotonated to become negatively charged. The second system, called the CYS system,
is the zinc-finger in which the cysteine amino acids remained in their neutral uncharged state.
The experiment X$-$ray crystal structure will be used as the initial structure of both systems.
Our results show that the overcharged zinc-finger is important for the
stability of the protein structure even in their binding to negatively charged DNA molecule.
In addition, the overcharged zinc-finger also has stable DNA binding pose whereas
 the complex deviates significantly from the experimental structure for the
 undercharged zinc-finger.
There are also fewer differences between the two protomers in this weak DNA-binding system.
Therefore, the main differences between the structures of sequence$-$identical protomers A and B
are due to their interactions with DNA.

This paper is organized as follows. After the introduction in Section 1, the detail of the computational procedure is 
presented in Section 2, the results are presented and discussed in Section 3. We conclude in Section 4. 

\section{Methods}

\subsection{Preparation of the simulation systems}
The structure of the PSA protein's zinc fingers and the DNA segment it binds to is downloaded from 
the Protein Data Bank (\url{https://www.rcsb.org/}) with PDB code 1R4I. This structure was resolved using
 X-ray crystallography method with a resolution of 3.1\r{A}\cite{shaffer2004}. 
The complex contains a DNA segment and two protein chains called protomers A and B, and four zinc ions. 
On each protein chain, the Cys542, Cys545, Cys559, and Cys562 amino acids bind to the first zinc ion (Zn$_{1}$) 
and the Cys578, Cys584, Cys594, and Cys597 amino acids bind to the second zinc ion (Zn$_{2}$)
 in a tetrahedral structure (Fig. \ref{fig:system}). Four zinc fingers are present on this complex.
Two zinc fingers are found on each protomer.
Then, we manually remove the hydrogen atoms from the thiol group of 16 zinc-binding cysteine amino acids
to investigate the difference between the CYS complex with cysteines in their natural state and 
the CYN complex with cysteines in the deprotonated state.

\subsection{Molecular dynamics simulation}

The periodic boundary condition is used in our simulation. After setting up the coordinates of the atoms, 
the periodic simulation box size is chosen such that the protomers and DNA complex on neighboring
periodic boxes are at least 3nm apart, which is substantially larger than the screening length of the solution 
(about 0.7nm at 150mM NaCl salt concentration). 
This distance is large enough to eliminate the finite size effect due to the long-range electrostatic interactions, 
and yet maintain a 
small enough system for the simulation to run within a reasonable period of time with the available computing resource. 
The systems are then solvated with water molecules in an explicit solvent simulation.
After solvation, Na$^{+}$ and Cl$^{-}$ ions are added to the system at the physiological 
concentration of 150mM by randomly replacing water molecules with ions. 
The total charge of the system is zero to maintain the neutrality. 
The systems are then subjected to an energy minimization procedure using the steepest descent method to 
remove potentially high energy contacts and overlapping atoms 
before performing the molecular dynamics simulation.

All-atom molecular dynamics simulation with explicit solvent model is carried out in this work.
The forcefield AMBER 99-ILDN\cite{showalter2007validation} is used to parameterize the protein molecules. 
The state of the art forcefield, PARMBSC1 \cite{Ivani2016} is used to parameterize the DNA molecule. 
Water molecules are parameterized using the TIP3P forcefield \cite{price2004modified}, a common and highly compatible forcefield
 for the chosen Amber forcefields. The GROMACS version 2018.3 software package\cite{hess2008gromacs}
 is used for the molecular dynamics simulation of the systems. 
 Each system is subjected to equilibration in NVT and NPT ensembles
 at the temperature of 298 K and at the pressure of 1 atm for 100 ns. 
Then, a long production run of 1000 ns each is used to take statistics. 
Nose-Hoover thermostat \cite{nose1984molecular,hoover1985canonical} is used to maintain the temperature of the systems. 
Parrinello-Rahman barostat \cite{parrinello1980m, parrinello1981polymorphic} is used to maintain the pressure of the systems. 
Both electrostatics and van de Waals interactions are cut off at 1.2nm. 
The long-range part of the electrostatic interactions among charges is calculated in the reciprocal $k$-space 
using the Ewald summation  via Particle Mesh Ewald method\cite{darden1993particle} at the fourth order interpolation. 
The long-range part of the van de Waals interactions among atoms is approximated as appropriate corrections 
to the energy and pressure. All covalent bonds are constrained using the LINear Constraint Solver (LINCS) algorithm
in order to increase the simulation time step to 2.5 fs \cite{hess1997lincs}. 

\subsection{Analysis the results of molecular dynamics simulation}
Analysis of the simulation results is performed using the corresponding tools 
provided in the GROMACS package, 
such as the root mean square deviation (RMSD) and 
the root mean square fluctuation (RMSF) 
for the backbone atoms of both protomers and the DNA on each upstream or downstream complex. 
The visualization of the 3D structures of the systems is performed using  
the program VMD version 1.9.3 \cite{humphrey1996vmd}. 
Some in$-$house python scripts are used for various tasks and
for combining different analysis softwares for RMSD-based clustering,
covariance matrix calculations, and principal component analyses.

\section{Results and discussions}

\subsection{Deviations and fluctuations of the structural backbone atoms of proteins and of DNA}

\begin{figure*}
	\centering
	\subfigure[]{\includegraphics[width=0.48\textwidth]{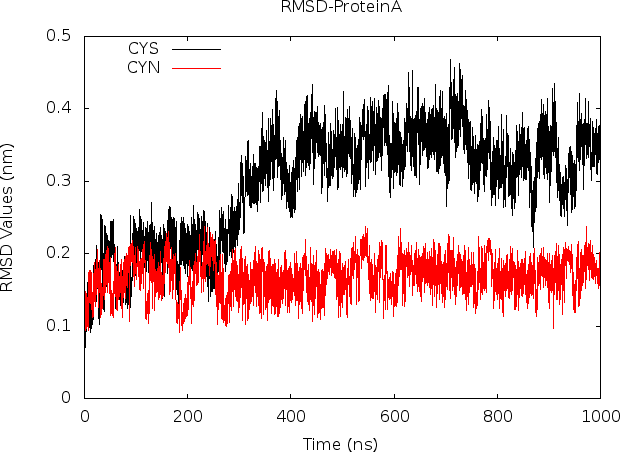}
	\label{subfig: dpa}}
	\subfigure[]{\includegraphics[width=0.48\textwidth]{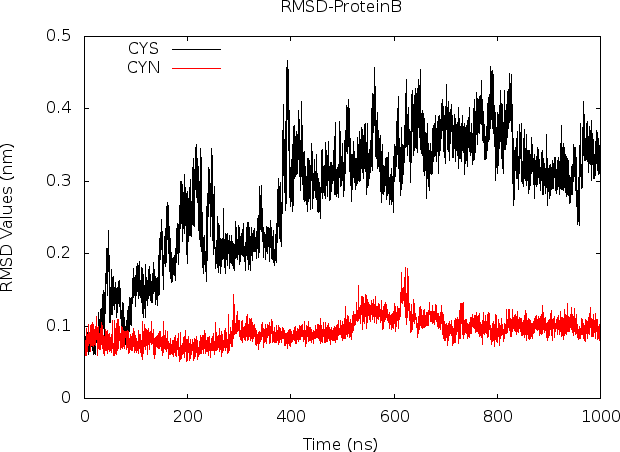}
	\label{subfig: dpb}}
	\subfigure[]{\includegraphics[width=0.48\textwidth]{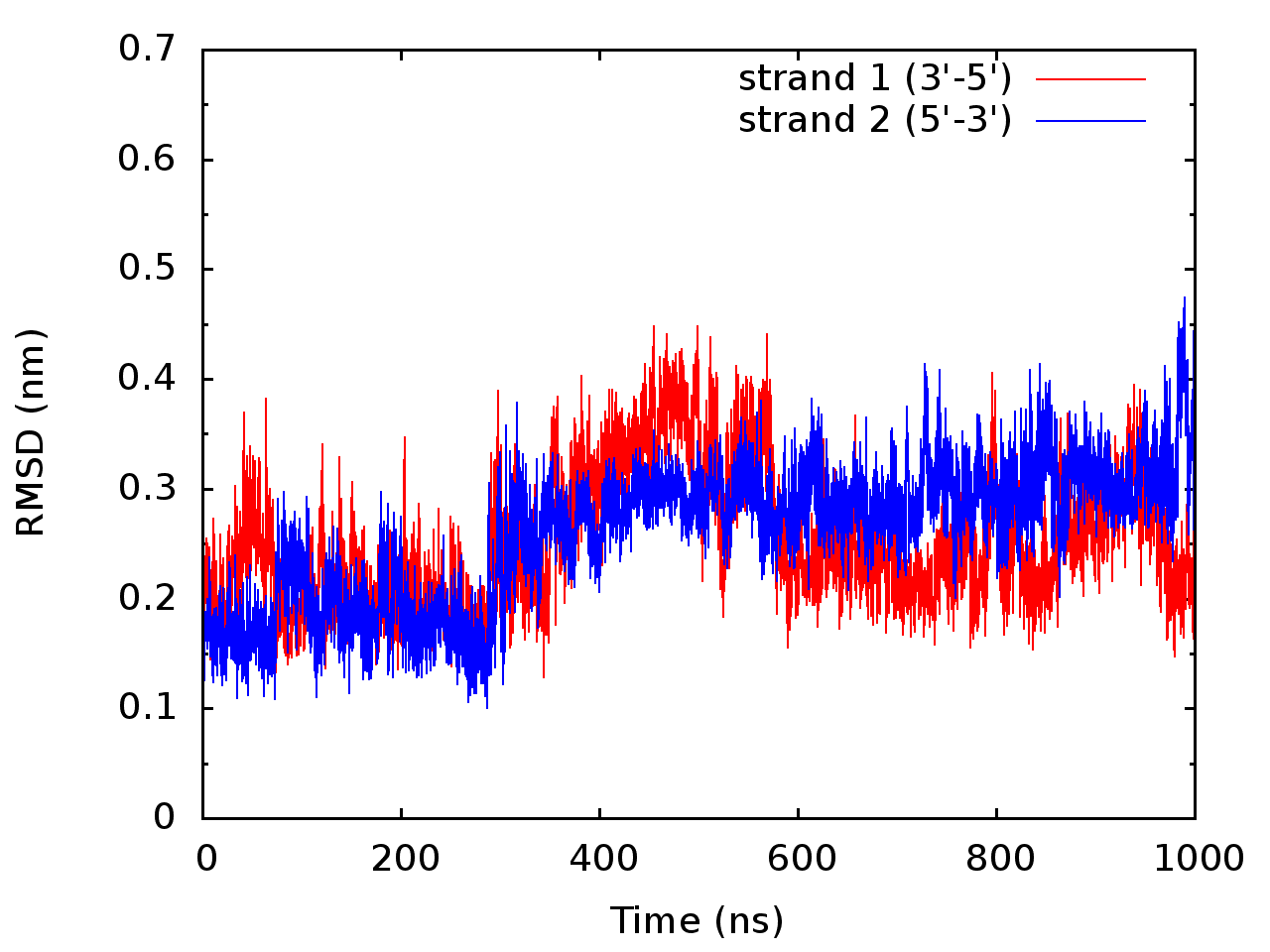}
	\label{subfig:ddn}}
	\subfigure[]{\includegraphics[width=0.48\textwidth]{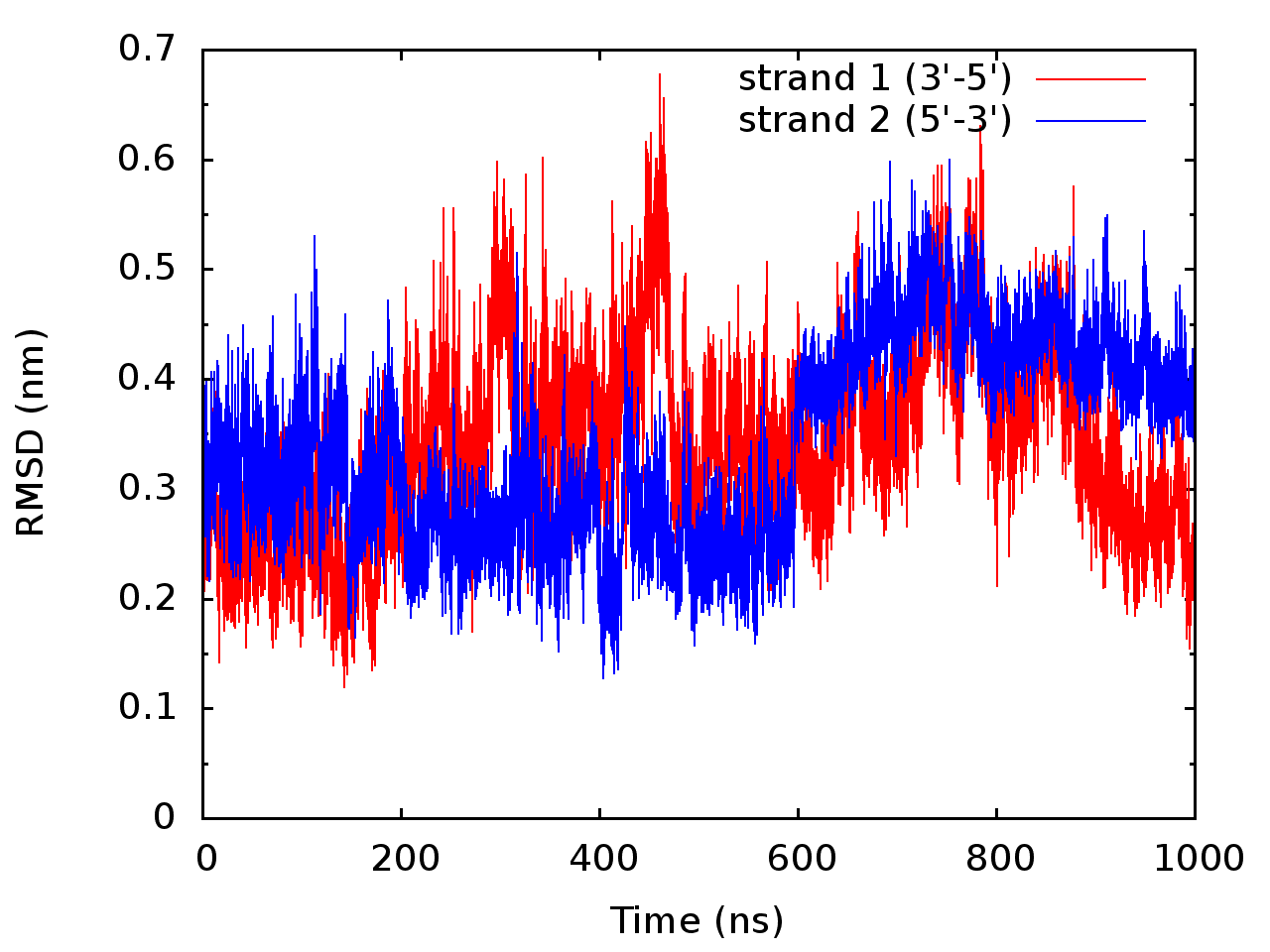}
	\label{subfig:dds}}
	\caption{The root mean square deviation (RMSD) of the backbone of the two protein chains A (a) and B (b) and 
the two complementary strands of the DNA molecules (c) and (d) 
from their experimental X$-$ray crystal structure as a function of time. 
For each figure, the CYN (deprotonated cysteine residues) and CYS (natural cysteine residues) are plotted for comparison. 
The stability of the CYN system is clearly demonstrated as its typical deviation (red line) 
is less than 2\AA\ from the native structure,
whereas the CYS system (black curve) deviates by 4\AA.  
Standard deviation for the DNA molecule is the same on both systems.}
	\label{fig:rmsd}
\end{figure*}

As a standard procedure, the initial analysis of the systems is performed by calculating the root mean square
deviation (RMSD) of the proteins from its native crystallized X$-$ray experiment structure. 
The backbone C$_\alpha$ atoms are used for the calculation
of the RMSD of the proteins. The O4' atoms (in standard deoxyribose nucleic acid nomenclature)
 of the sugar group of the backbone of the DNA strand are used
 to calculate the RMSD of nucleic acids. The results are plotted in Fig. \ref{fig:rmsd}. 

The deprotonated CYN state (red line) clearly shows
higher stability with only 2\AA\ deviation from the native structure for protomer A and 1\AA\ deviation
for protomer B. The deviations for the same protein chains in the CYS system (black line) have a much
higher value of up to 4\AA. The CYN complex is stable
throughout the simulation run  (with protomer B being more stable than protomer A), 
whereas the RMSD values for the CYS system reach their plateaus 
only after about 500 ns for both protein chains. 
We will see later that this is related to the
reorganization of the secondary structures, the changes in the unstable binding pose
of the CYS proteins to the DNA, and the zinc ions move to different locations with
lower electrostatic energy. As a result of this RMSD analysis, 
only configurations from 500 ns onward are processed in all later statistical analysis
of the reference structures of the CYS system.
It should be stressed that, from the point of view of computer simulation, 
such plateauing behavior means that new equilibrium has been established. 
Of course, one cannot conclude for certain that this change is irreversible with only 1000ns of simulated time. 
However, considering that Zn$^{2+}$ is a high valence ion 
with strong electrostatics interactions, we believe that this change in the CYS system is
irreversible and not transient, at least for the locations of the zinc ions.

The RMSD deviation for the DNA molecule is plotted in Fig. \ref{subfig:ddn} for the CYN system 
and in Fig. \ref{subfig:dds} for the CYS system. Unlike the deviation of protein structures,
the RMSD plotted for DNA strands are similar in both systems. Although
for CYS system, deviation as large as 7\AA\ are observed, and seemed to coincide
with a large deviation in protomer B as it also deviates strongly at around 400ns. 
DNA RMSD in both systems show a plateau after 400ns, and settle at a saturated value of 4\AA\ deviation
as the DNA molecule equilibrates its binding pose to the protein chains.
This value is the same for both complimentary strands of the DNA, thereby suggesting
that the two strands always remain in a double helix state and move
together. This stresses the structural stability of the DNA double helices, 
unaffected by the change in protein configuration. 

\begin{figure*}
	\centering
	\subfigure[]{\includegraphics[width=0.48\textwidth]{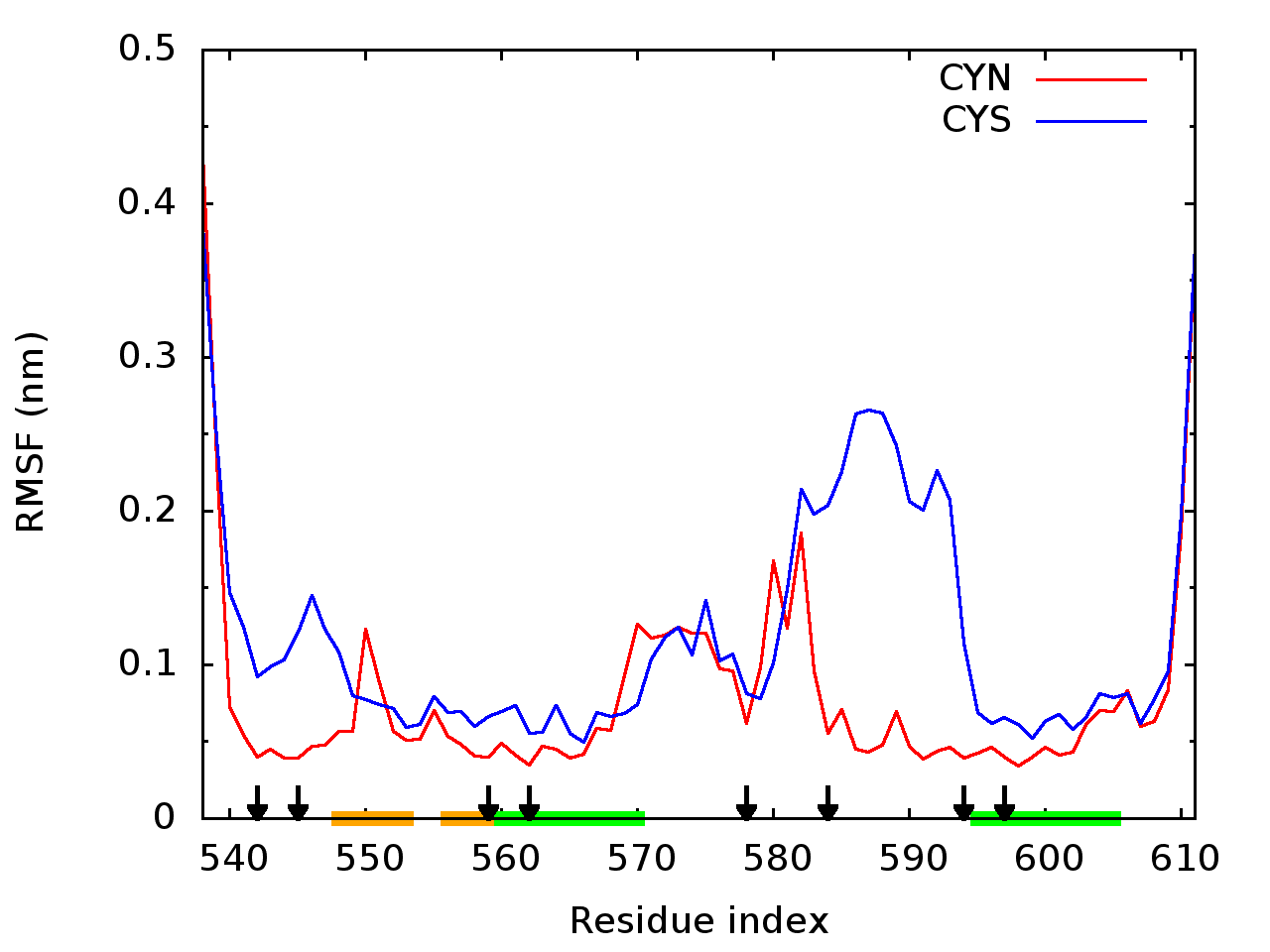}
		\label{subfig:fpa}}
	\subfigure[]{\includegraphics[width=0.48\textwidth]{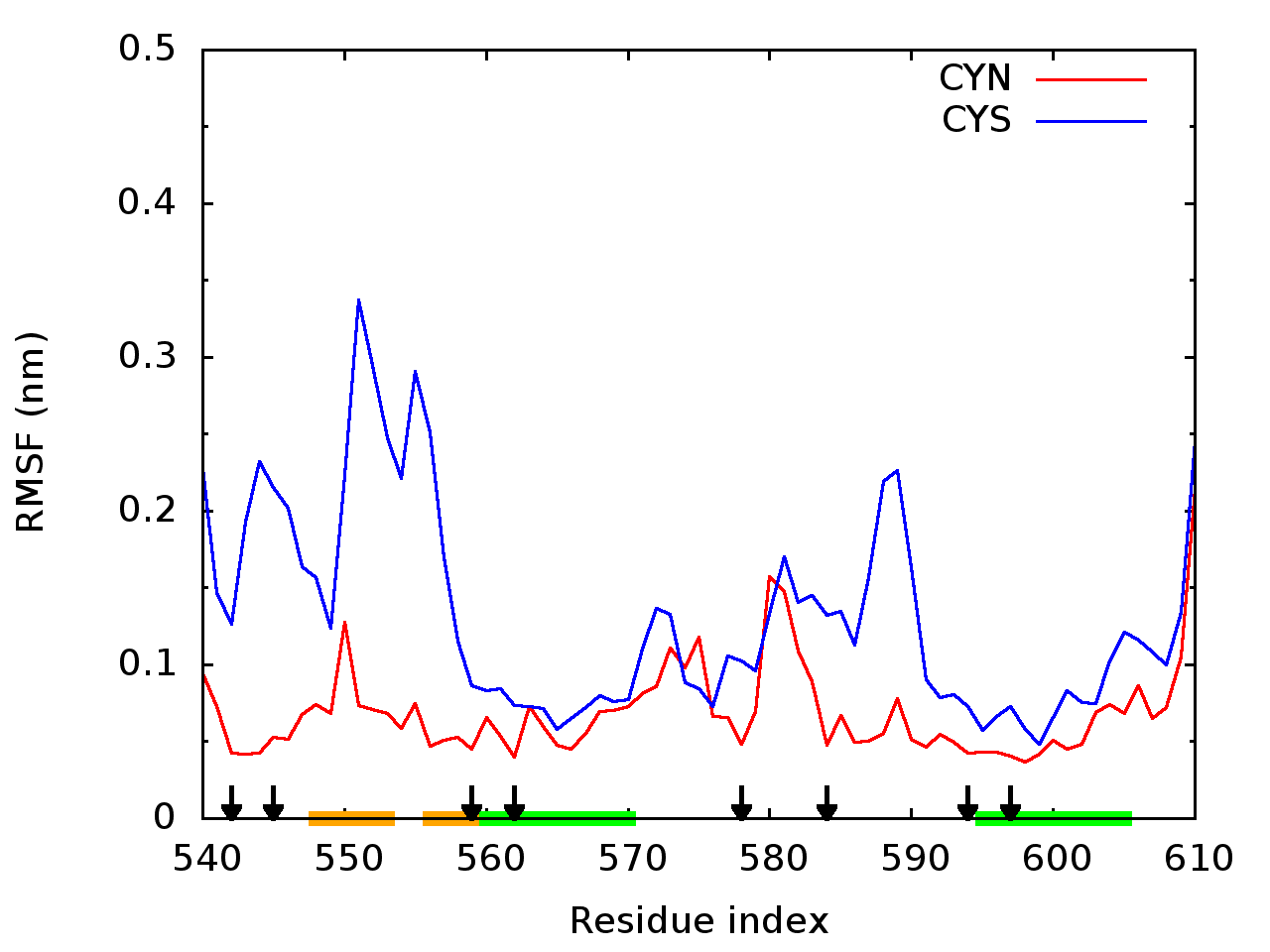}
		\label{subfig:fpb}}
	\subfigure[]{\includegraphics[width=0.48\textwidth]{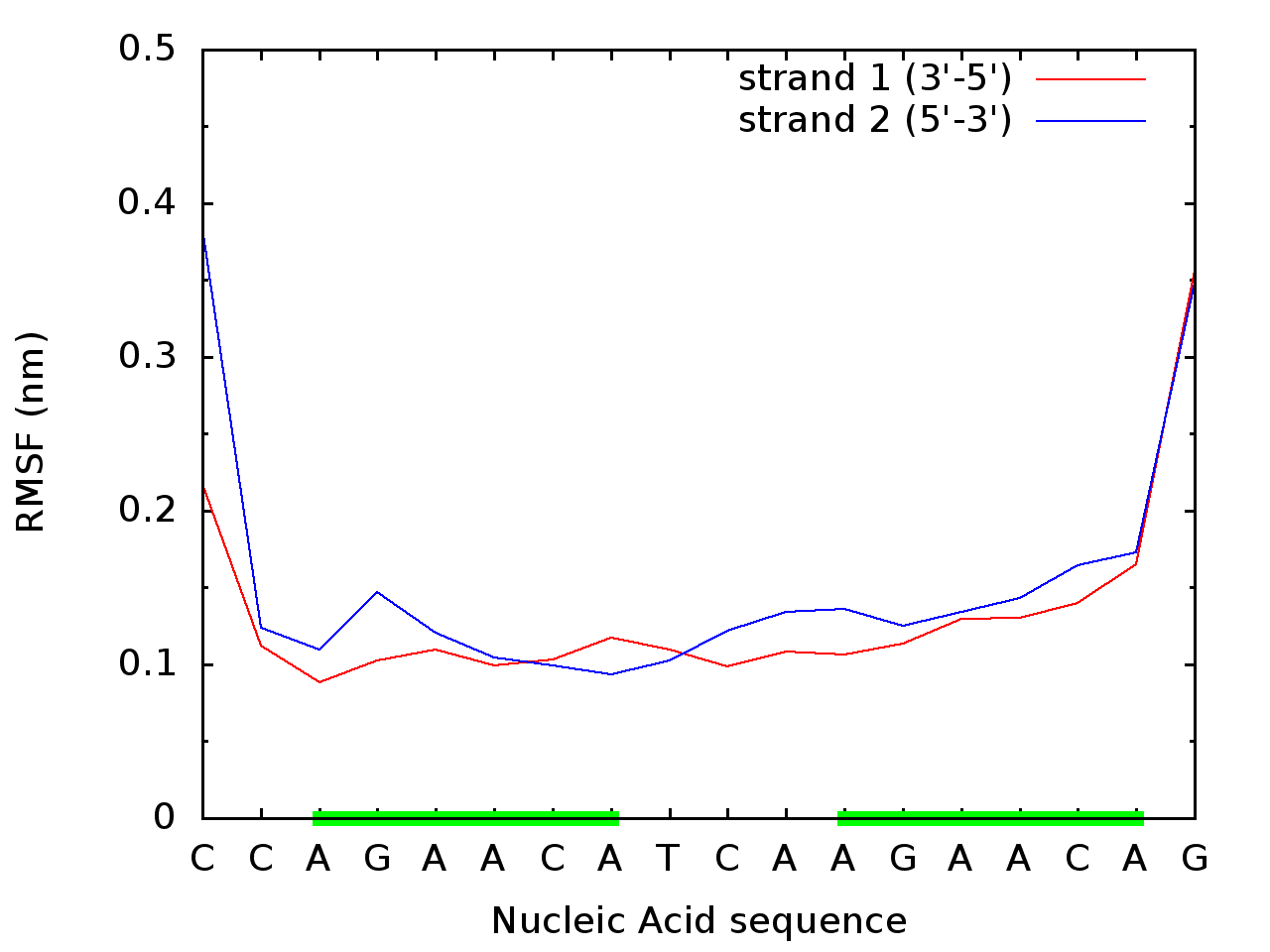}
		\label{subfig:fdn}}
	\subfigure[]{\includegraphics[width=0.48\textwidth]{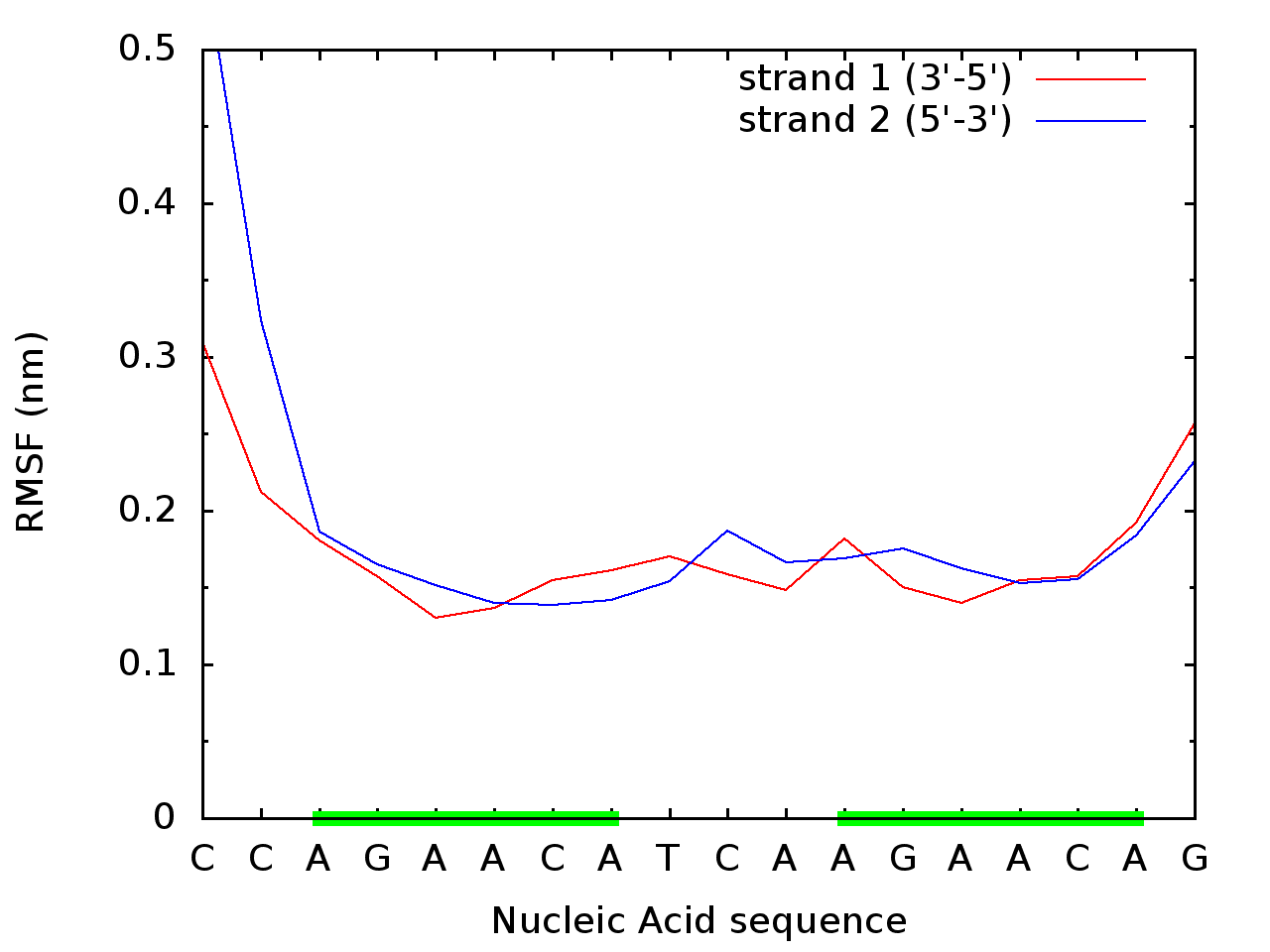}
		\label{subfig:fds}}\vspace{-0.16cm}
	\caption{The root mean square fluctuation (RMSF) of backbone C$\alpha$ atoms of protein chains A (a) and B (b).
The RMSF of the backbone atoms O4' of the two complementary strands of the DNA molecule for CYN system (c) 
and CYS system (d). For sub$-$figures a) and b), 
the green bars on the x$-$axis are the location of $\alpha-$helix residues, and the yellow bars 
are the location of $\beta-$sheet residues. The downward arrows on these axes are the locations of the
eight cysteine residues involved in binding of the two Zn$^{2+}$ ions of the zinc-fingers. 
For sub-figures c) and d), the green bars 
are the location of the upstream and downstream nucleic acid sequences that the zinc$-$figure proteins bind to.
}
	\label{fig:rmsf}
\end{figure*}
Next, we calculate the time-averaged RMSF of the atoms of the protomers and DNA 
backbone residues. 
RMSF is directly related to how the deprotonated state of Zn$-$Cys4 complex can affect the structural rigidity
of the molecules. Once again, only atomic fluctuations of the C$_\alpha$ atoms of the protein
and of the O4' atoms of the DNA are considered, because these backbone atoms are representative of the
overall structure of the molecules more than the side chain atoms. The results of the atomic
fluctuations for the CYN and CYS systems are shown in Fig. \ref{fig:rmsf}.
Figs. \ref{subfig:fpa} and \ref{subfig:fpb} show that
the average fluctuation values of the C$\alpha$ backbone atom of each amino acid residue
are almost always smaller for the CYN systems (around 0.5 \AA) compared with the CYS
system (1 \AA\ to 3 \AA). This is especially true for most of the four cysteine residues that make up
the second zinc-finger of protomer A and the first zinc-finger of protomer B. As
we will see later, this difference in fluctuation is due to the zinc ions of these zinc-fingers
being pulled into the aqueous solution in the CYS system in its new equilibrium.
This confirms that deprotonated, negatively charged cysteine residues stabilize 
the zinc-finger structure even in the presence of negatively charged DNA molecule. 

Another observation is the large fluctuations of the
$\beta\beta\alpha$ zinc$-$finger that binds to the major groove of the DNA in protomer B (from residue GLU548 
to residue ALA570). As we will see later, this zinc finger structure is disrupted strongly by the protonation state of 
the cysteines. The region that binds both the protein chains to the DNA are very stable in CYN system
with a fluctuation of only 0.5 \AA. 
This again confirms that deprotonated cysteines stabilize both the zinc finger structures,
and the DNA-binding pose of zinc-fingers, 
even if both DNA and zinc-fingers are negatively charged in the CYN system.
The RMSF value of 0.5\AA\ is remarkably lower than the typical 5\AA\ RMSF value
for regular folded protein in solution. This means that DNA binding stabilizes the protein structure
of these zinc fingers.

Figs. \ref{subfig:fdn} and \ref{subfig:fds}) show the atomic fluctuation along the backbone of the
nucleic acid segment. Both strands show very similar values, 
dominantly in the range of 1 \AA\ to 2 \AA\ (excluding the free moving end of each strand) and only
very minor difference between CYN and CYS systems. The two upstream and downstream sequence backbones 
(the green bars in the x$-$axis) behave similarly and stably for both strands. 
The results show the same trend as that in the RMSD analysis. The structural
rigidity of the DNA double helix is weakly affected by the deprotonated state of the binding proteins.

\subsection{Disruption to the secondary structure of the zinc$-$fingers}

\begin{figure*}
	\centering
	\subfigure[]{\includegraphics[width=0.98\textwidth]{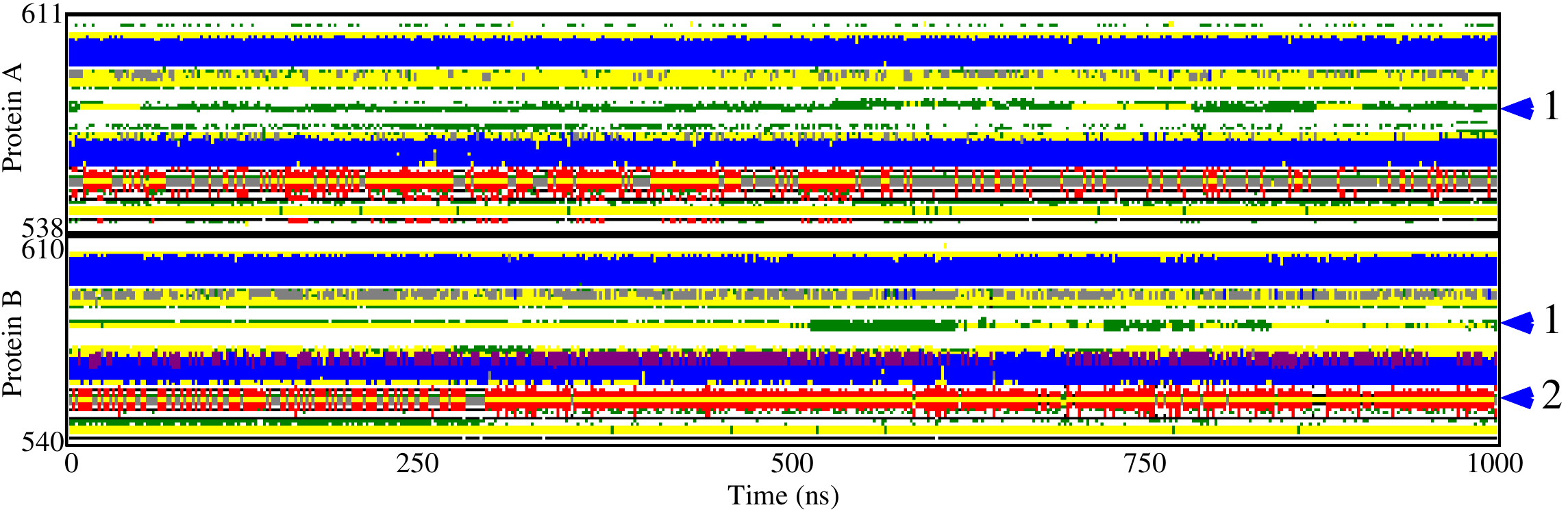}
		\label{subfig:dsspCym}}
	\subfigure[]{\includegraphics[width=0.98\textwidth]{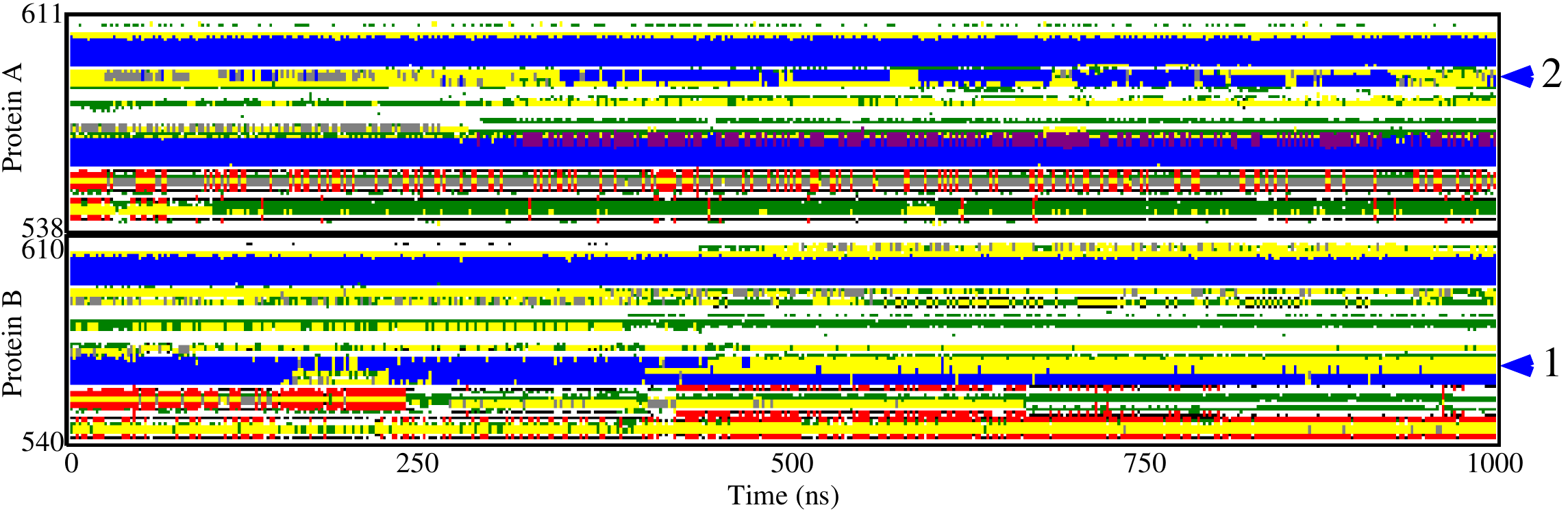}
		\label{subfig:dsspCys}}
	\caption{The secondary structures of zinc$-$finger proteins as a function of time for CYN system (a)
and CYS system (b). The vertical axis is the residue index.
The DSSP classification system is used, and the residues are color coded
as follows: blue, $\alpha-$helix; gray, $3-$helix; violet, $5-$helix; red, $\beta-$sheet; black, $\beta-$bridge; 
yellow, hydrogen bonded turn; green, bend; and white, coil. The main helices and $\beta-$sheets
of the CYN system
are stable for the whole simulation length, wherease the CYS system shows major disruptions to secondary structures
 from about 400 ns onward. The arrows on the right of (b) show residue locations 
 where these major disruptions occur compared with (a). See text for more discussion.
The arrows on the right of (a) are used in discussion of Fig. \ref{fig:rmsdclusterCYN}. 
	}
	\label{fig:dssp}
\end{figure*}

Let us analyze how the secondary structures
of the proteins are affected by the protonation state of these zinc finger amino acids. In Fig. \ref{fig:dssp},
the changes in secondary structure during simulation are shown in the top figure 
for the two protomers of the CYN system, and in the bottom figure
the two protomers of the CYS system. 
The definition of the secondary structure follows
the standard DSSP classification system. The major $\alpha-$helices involved in the zinc$-$fingers are shown in blue.
The helix from the residues GLU560 to ALA570 sits at the DNA's major groove, 
whereas the other helix from the residues PRO595 to ALA605 aligns along the DNA's principle axis. 
Comparing the change in the secondary structures of the proteins overtime for the CYN and CYS systems 
immediately reveals a major disruption around 400ns in the CYS system, as already inferred from the RMSD analysis.
From these figures, the effect of zinc-binding in CYS neutral state
influences the secondary structure of zinc$-$finger protein differently for the upstream versus
the downstream binding configurations. 
For the downstream binding complex (protomer B),
the first zinc finger is affected more than the second zinc-finger. 
Specifically, the $\alpha-$helix from residues GLU560-ALA570 
melts and shorten by half from 400ns onward (blue arrow `1' in Fig. \ref{fig:dssp}b). 
In later analysis, we will be able to see that in the CYS system, the Zn$^{2+}$ ion 
unbinds from the cysteines and moves to bind with the negatively charged DNA phosphate backbone atoms instead.
The shorten helix, however, remains bound with DNA and only disorients inside the major groove, 
leading to higher fluctuations and deviations.

For the upstream binding complex (protomer A), the second zinc$-$finger associated with 
the second helix more affected than the first zinc-finger. This helix from residues PRO595 to ALA605 of protomer A 
in the CYS system shows transient extension to include more residues 
during the time frame of 400 - 900 ns (blue arrow `2' in Fig. \ref{fig:dssp}b). 
This transient event is due to this second zinc ions turn away from DNA to face 
the solvent molecule and detach from the $\alpha-$helix. This results
in the helix temporarily recruits more amino acids onto itself.

On the contrary, Fig. \ref{subfig:dsspCym} for the CYN system clearly shows that zinc-ion-overcharged state
is important. Both helices of the zinc$-$fingers for both upstream and downstream binding complex remains
stable during the whole simulation time of 1 microsecond. Only the unstructured loops (blue arrows `1'
in Fig. \ref{fig:dssp}a) show large significant changes during simulation, which is natural for such flexible regions. 
Additionally, the native $\beta-$sheets of the proteins recovers transiently as shown in blue arrow `2' in
Fig. \ref{fig:dssp}a). See the CYN configuration discussion near Fig. \ref{fig:rmsdclusterCYN}
for more details on this secondary structure.

\begin{figure}
	\centering
	\subfigure[]{\includegraphics[width=0.48\textwidth]{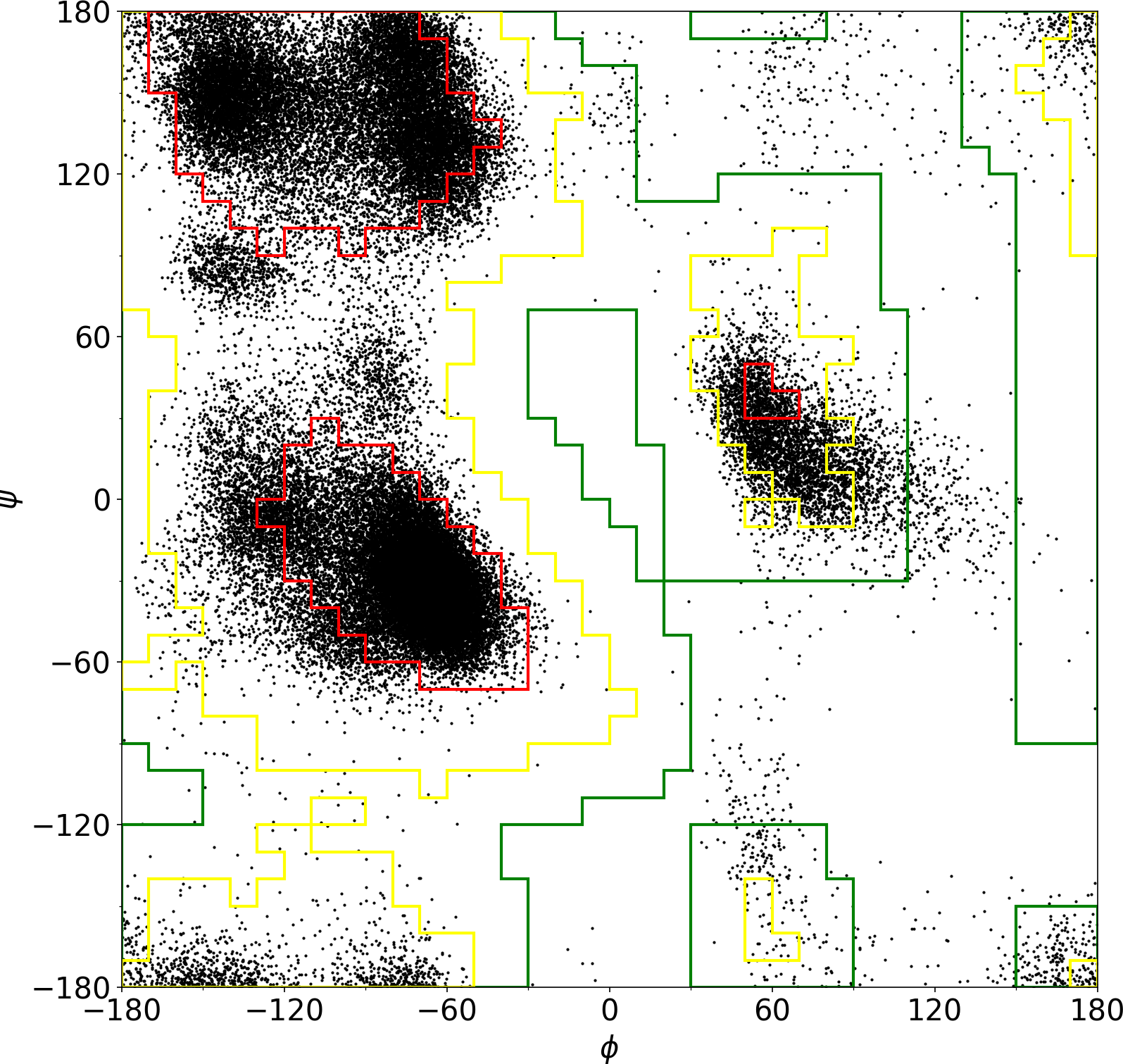}
		\label{subfig:ramaCym}}
	\subfigure[]{\includegraphics[width=0.48\textwidth]{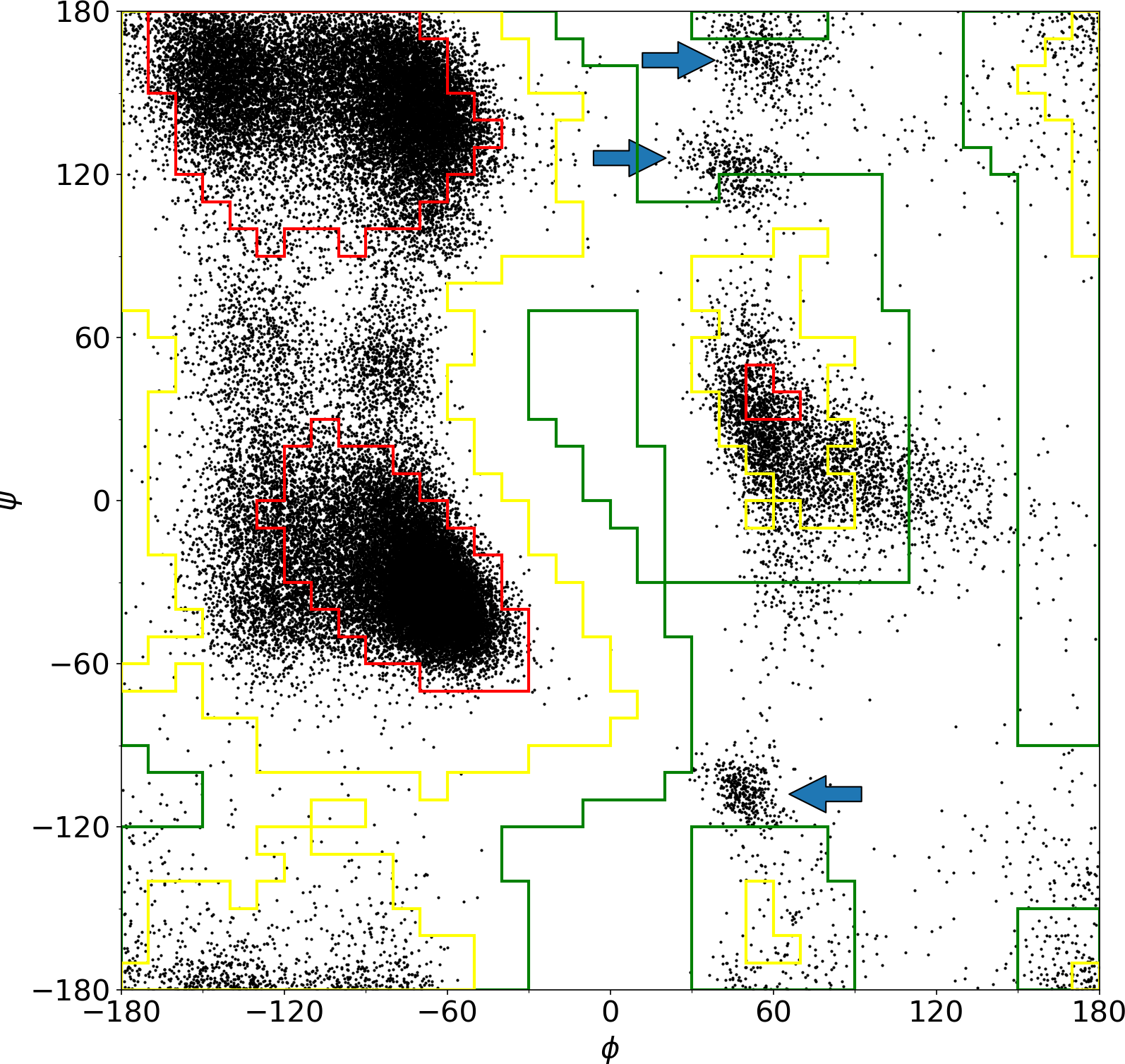}
		\label{subfig:ramaCys}}
	\caption{The Ramachandran plot of the proteins in deprotonated CYN system (a) and neutral CYS system (b).
The "red" zone is the favorable region where the structures of $\beta-$sheets 
 and $\alpha-$helices are located, the "yellow" zone is the allowed region, 
 and the "green" zone is the generously allowed region. 
The CYN system lies mostly within the allowed region. 
The CYS system shows high instabilities 
with many pairs $(\psi, \phi)$ in the 'unfavorable' high energy zone. 
The blue arrows in the plot for the CYS system
show the regions of the major accumulations of pairs in the `unfavorable' zone.
See text for discussion.
	}
	\label{fig:rama}
\end{figure}

Another measure of the stability of the structure of the proteins in these systems is to calculate 
the Ramachandran plot for the angles of the C$_\alpha$ backbone atoms of the peptide chains. 
The results for the two systems are plotted in Fig. \ref{fig:rama}.
The regions of favorable values of the two dihedral angles ($\psi$, $\phi$)
of the proteins' peptide backbone are outlined in red for clarity.
Most of the values for the proteins of the two systems
indeed expectedly fall inside these red regions. 
Additionally, the "yellow" and "green" regions are for the "allowed" and 
"generously allowed" values, respectively. 
The "unfavorable" region with high energy cost for these values of the angle pairs
are outside the green boundary. This plot shows that the neutral CYS system 
has many high energy angle pairs. Interestingly,
pairs accumulate in certain regions in this
``unfavorable'' zone, as shown by the blue arrows in Fig. \ref{fig:rama}b.
 
 A calculation of the dihedral pairs $(\phi, \psi)$ for individual residues (not shown)
in the CYS system indicates that these accumulation regions is due mostly to
the dihedral angles of the three residues, SER580, ASN582 and ASP583. These  amino acids 
are polar or charged residues in the loop between the first and second cysteines, CYS578 and CYS584,
of the second zinc-finger of protomer B.
Later configuration cluster analysis (Fig. \ref{fig:rmsdclusterCYS}c and \ref{fig:rmsdclusterCYS}d) shows that
the zinc ion of this zinc-finger in the CYS system
leaves the Cys4 pocket and moves toward the aqueous solution to
lower its electrostatic energy. The electrostatic interaction of the positive +2e zinc ion to the
oxygen atoms on these polar and charged amino acids are so strong 
that the zinc ion pulls (and fixes) the position of these atoms.
This pull causes the dihedral angle pairs in these amino acids to fall
into the `unfavorable' zone and create the three accumulation regions observed.

Unlike the CYS system, the overcharged deprotonated CYN system avoids
these high energy regions and is mostly compact in the allowed regions (Fig. \ref{fig:rama}a).
This result confirms the stability of the overcharged configuration of CYN in the DNA$-$binding complex.

\subsection{Hydrogen bonding stability}

\begin{figure}
	\centering
	\subfigure[]{\includegraphics[width=0.45\textwidth]{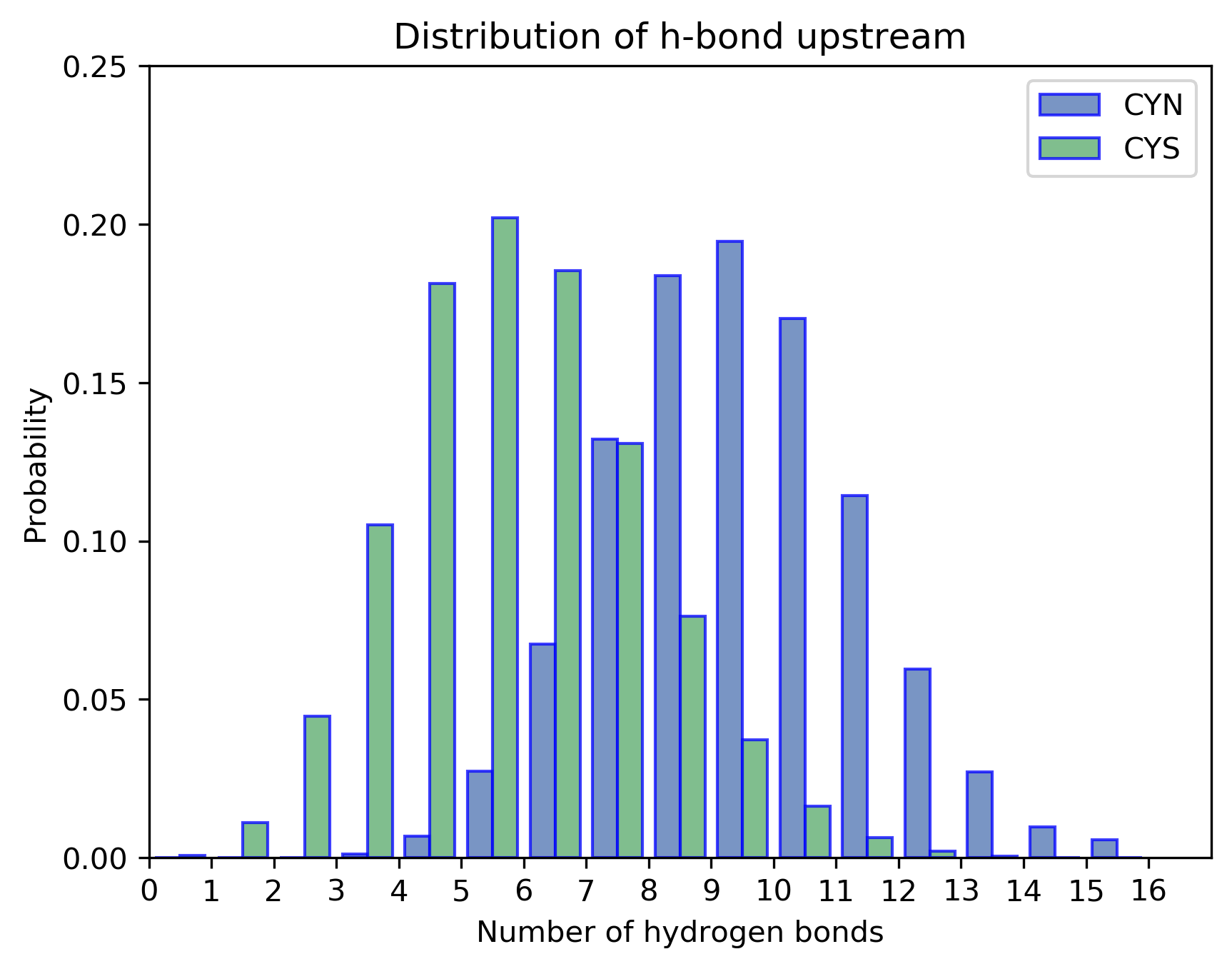}
		\label{subfig:hban}}
	\subfigure[]{\includegraphics[width=0.45\textwidth]{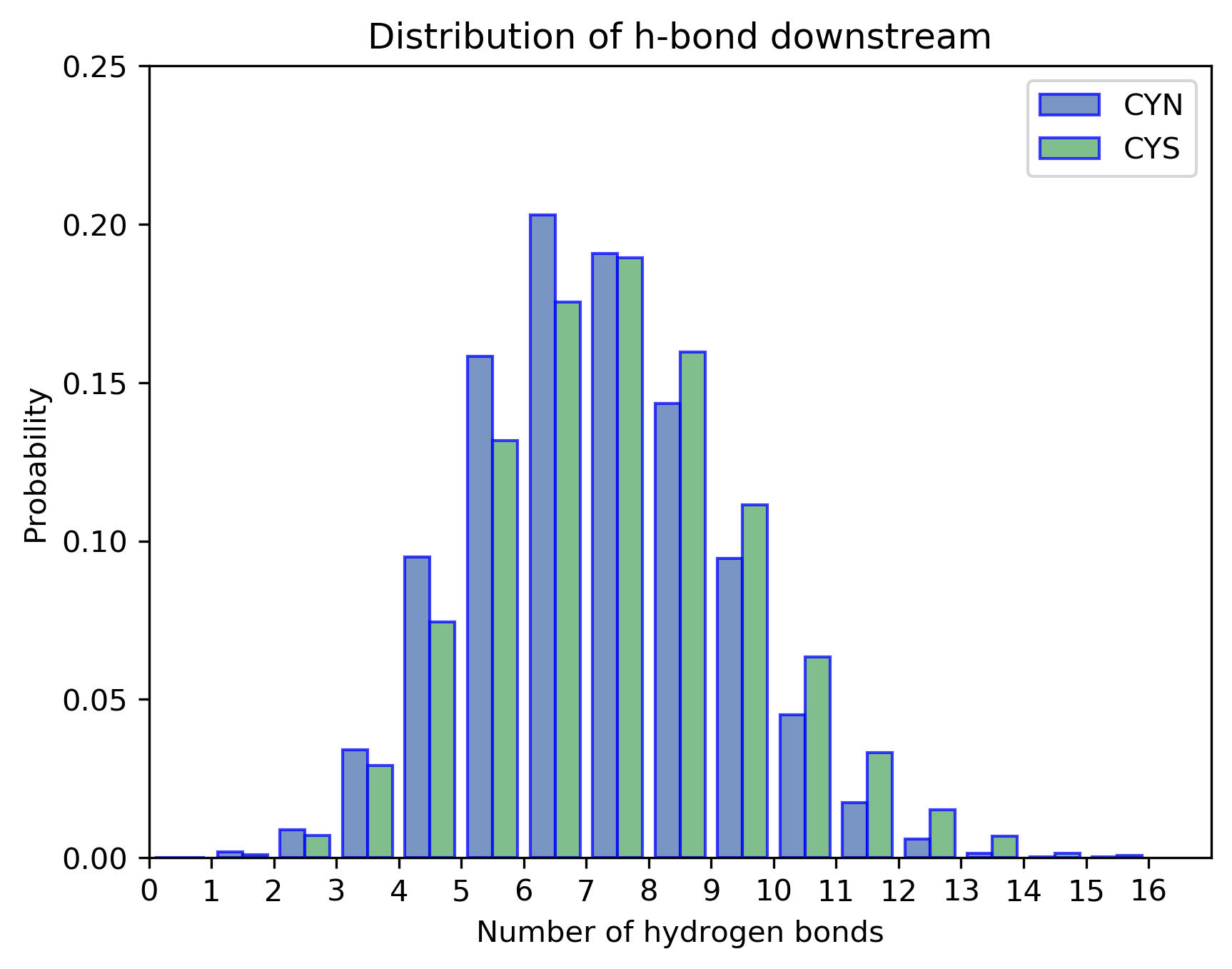}
		\label{subfig:hbas}}
	\caption{The distribution of the number of hydrogen bonds between the upstream nucleic sequence with protomer A (a)
and downstream nucleic sequence with protomer B (b) as calculated through the VMD program. 
In each sub$-$figure, the distribution
for the CYN system is shown in blue color and the CYS system is shown in green color.}
	\label{fig:hbond}
\end{figure}

Previous simulation works have shown that hydrogen bonds are unique in the presence of zinc ion binding \cite{Godwin_2017}.
The structure of the folded protein shows the narrowest distribution of hydrogen bonds in the overcharged state.
Therefore, one naturally asked how this state influences hydrogen bonding with the nucleic acids in their complexation
with the DNA molecule. The distribution of hydrogen bonds for protomers A and B
with the upstream and downstream DNA sequences are plotted in Fig. \ref{fig:hbond}. In each plot, 
the values for the overcharged CYN system are colored light blue and
 those for the undercharged CYS system are colored light green.  
One can see from this figure that the hydrogen bonds of protomer B with the downstream sequence
are stable in both systems. A Gaussian fit to these histograms gives a mean value of 6.58 bonds with a standard
deviation of 1.94 bonds for the protomer B of CYN system and a mean value of 7.03 bonds
with a standard deviation of 2.12 bonds for the protomer B of the CYS system. Thus,
the CYN system shows a slightly narrower distribution, indicating more unique bonding.

Protomer A on the other hand shows a loss of several hydrogen bonds 
in the undercharged CYS state. Gaussian fit to these histograms gives a mean value of 8.97 bonds 
with a standard deviation of 2.01 bonds for the protomer B of CYN system and a mean value of 5.39 bonds
with a standard deviation of 1.95 bonds for the protomer B of the CYS system.
The total number of hydrogen bonds lost in the CYS system is about 3.5 bonds, near a third
of the total. Later cluster analysis, which is used to investigate the representative structures,
reveals that this result is caused by the lifting of the first zinc finger further away 
from the DNA to push the zinc ion deeper into the aqueous solution. 
For protomer B, due to its dimeric binding to protomer A, this zinc finger 
slightly more stable in its binding with DNA. 
The second zinc finger maintains its hydrogen bonds in the CYS system for
both protomers A and B.

It should be noted that these hydrogen bonds are heuristically determined from each configuration
based on the geometric criteria that the distance between the donor and acceptor atoms
is less than 3.5 \AA, and the angle of the three atoms making up the hydrogen bond is less than 30$^{\circ}$. 
A full quantum mechanical calculation is needed to identify actual hydrogen bonds
among the protein and the DNA. However, such calculations  for such a large system are beyond 
our computing capability. For the purpose of this work,
the classical criteria are enough to show the differences between the CYS and CYN systems.

\subsection{RMSD-based clustering and simulated representative structures}

Let us now move to investigate important dynamical features of the zinc$-$finger DNA binding complex. As a first
step, we use RMSD-based clustering analysis to group configurations of the 1 $\mu$s trajectories into similar 
configurations. This procedure, coupled with the principal component analysis later, provides a detailed insight into 
the various macrostates of the binding complex, its collective motions, and potential kinetic traps.
 
In all results listed in this work, the RMSD cutoff value of 0.15 nm is used to distinguish neighboring configurations. 
This value is chosen through trial and error to find the most reasonable number of clusters of configurations. 
For a large cutoff value, all configurations are neighbors and only one cluster is generated. 
Vice versa, for small cutoff value,
there are too many clusters of configurations generated, which defeat our purpose. 
In fact, by varying this value and counting
the number of clusters of configurations generated, one identifies a cutoff value for which this number show a sharp rise
in the number of configurations and a decrease in the probability of the most populous and 
lowest free energy cluster. Ultimately, the value of 0.15 nm is chosen as the threshold cutoff.
The results of distributing all the trajectory configurations into clusters using this RMSD cutoff value
is shown in Table \ref{tbl:rmsd} for the two simulated systems.
\begin{center}
	\begin{table}[h]
	\small
		\centering
		\caption{\label{jfonts} RMSD-based clustering of the structures of the protomers. 
The number of different clusters of structures and the probabilities of the
three most populous lowest free energy clusters are shown.} 
		\begin{tabular}{@{}l*{15}{||cc|cc}}
			\hline
				&\multicolumn{2}{c}{Protomer A} &\multicolumn{2}{c}{Protomer B} \\
			\hline
				&CYN&CYS &CYN&CYS \\
			\hline
			Number of clusters		& 8 	& 39 & 2		& 32	\\
			\hline
			Probability of 1st  cluster		& 92.5\%  & 51.9\%		& 99.92\%		& 39.3\%	\\
			Probability of 2nd cluster		& 4.5\%	  & 10.4\%		&  0.08\%		& 24.5\%	\\
			Probability of 3rd cluster 	&  1.6\% &  8.9\%		&  0\% & 12.1 \%	\\
			Percentage unclustered	& 0.1\%     & 0.2\%    &  0\%       & 0.2\%\\
			\hline
		\end{tabular}
	\label{tbl:rmsd}
	\end{table}
\end{center}
Table \ref{tbl:rmsd} shows how the protomers in the CYS system are much
unstable and strongly fluctuating compared with those in the CYN system. For protomer A,
the CYN system has only eight distinct clusters, with the lowest energy cluster having almost 93\% probability. 
Configurations of protomer B in the same system could be distributed into only two clusters 
with the lowest free energy having near 100\% probability. These data show that
the CYN system is very stable and stays close to the experimental ground state X$-$ray structure. On the other hand,
in the CYS system, configurations of protomer A are distributed into 39 clusters, with the three lowest free energy
clusters occupying 70\% of the total time. In the same system, configurations of protomer B can be 
distributed into 32 clusters, with the three lowest free energy clusters occupying about 75\% of the time. 
In both cases, the binding of protomer B to DNA is stronger than protomer A, as previously mentioned.

\begin{figure*}
	\centering
	\subfigure[]{\includegraphics[width=3.5cm]{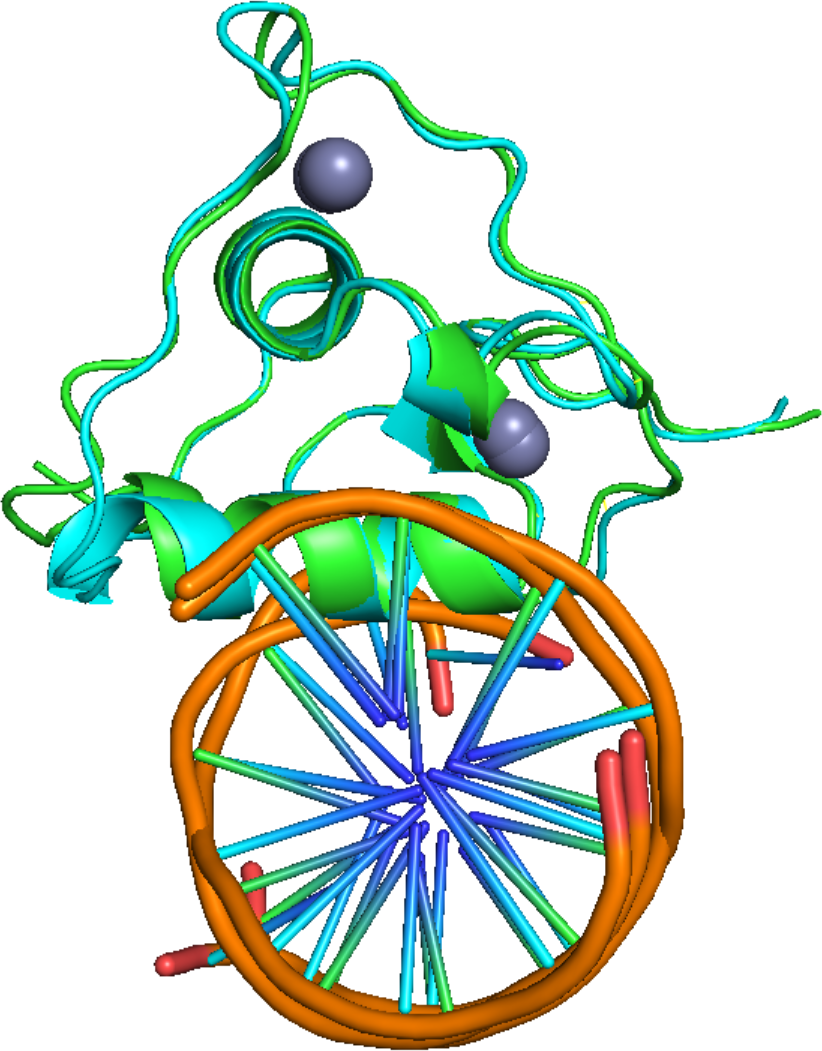}
		\label{subfig:rsmdCAN1}}
	\subfigure[]{\includegraphics[width=3.8cm]{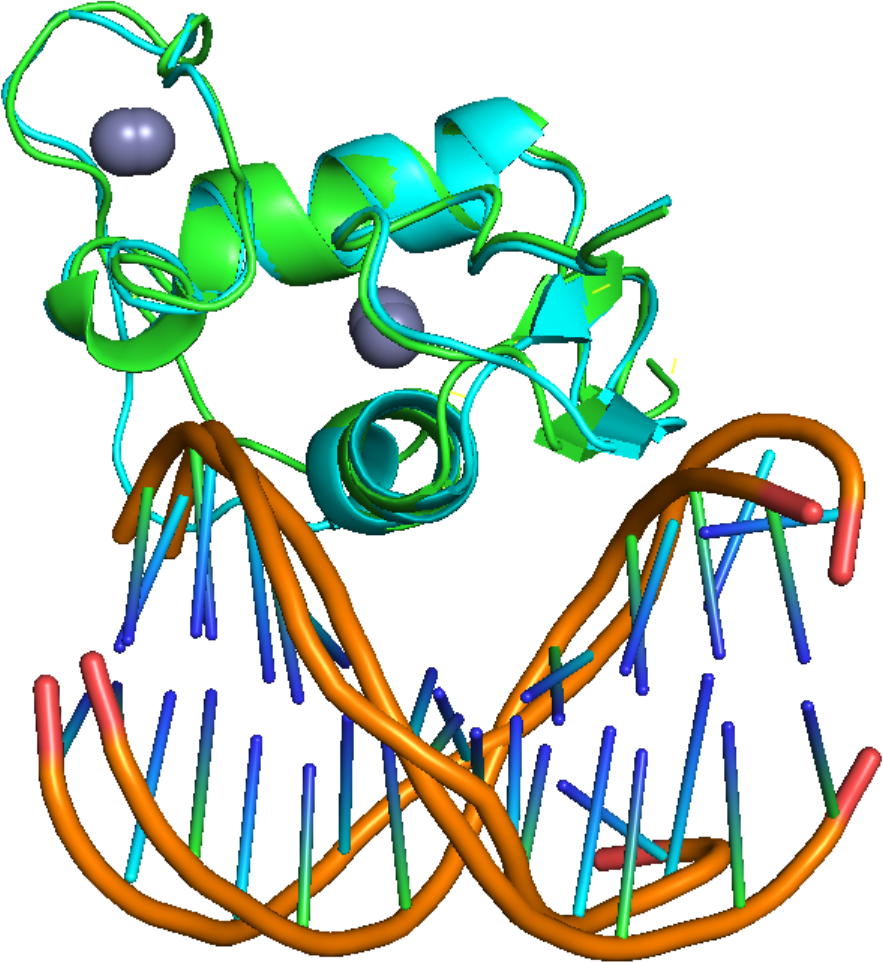}
		\label{subfig:rmsdCAN2}}
	\subfigure[]{\includegraphics[width=3cm]{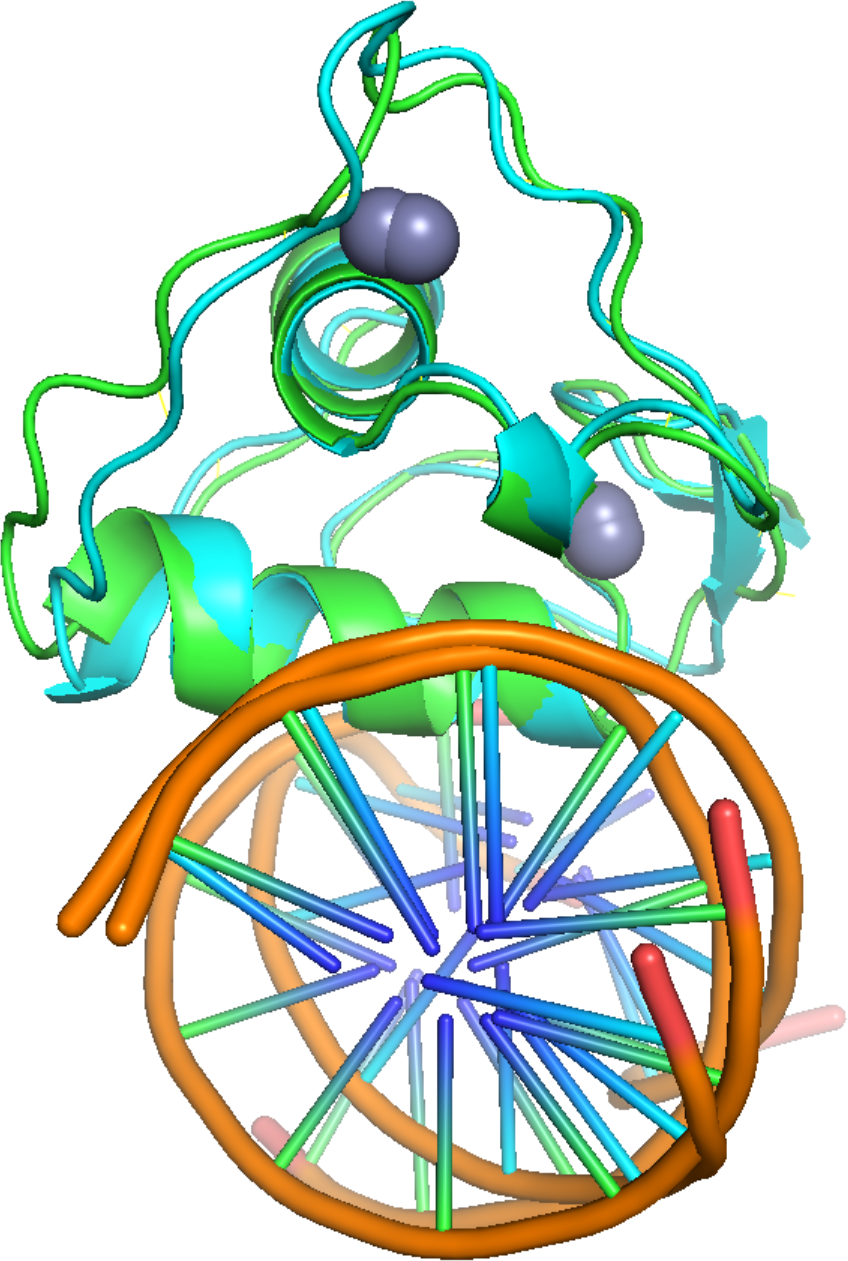}
		\label{subfig:rmsdCBN1}}
	\subfigure[]{\includegraphics[width=3.8cm]{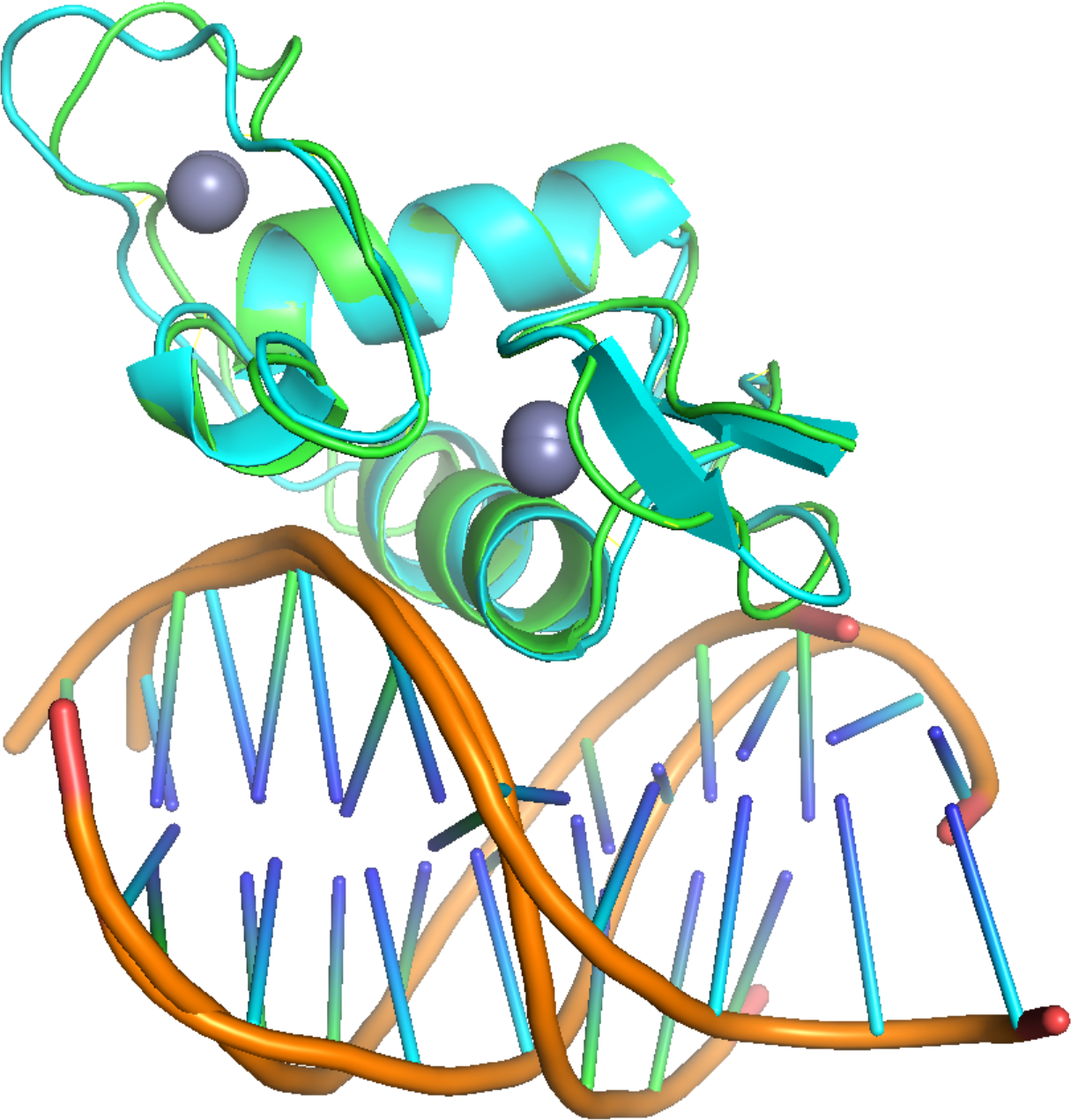}
		\label{subfig:rsmdCBN2}}
	\caption{Comparison of the central representative configuration of the dominant cluster 
for protomers A (a and b) and B (c and d). 
The experimental X$-$ray structure is shown in green, and 
the configuration obtained from the simulation is shown in cyan.
For each protomer, the top view (along the DNA axis) and side view are presented for clarity. 
The system is the overcharged CYN system. The simulated structure shows excellent agreement 
with experimental structure, stressing its stability.
	}
	\label{fig:rmsdclusterCYN}
\end{figure*}

\begin{figure*}
	\centering
\includegraphics[width=0.46\textwidth]{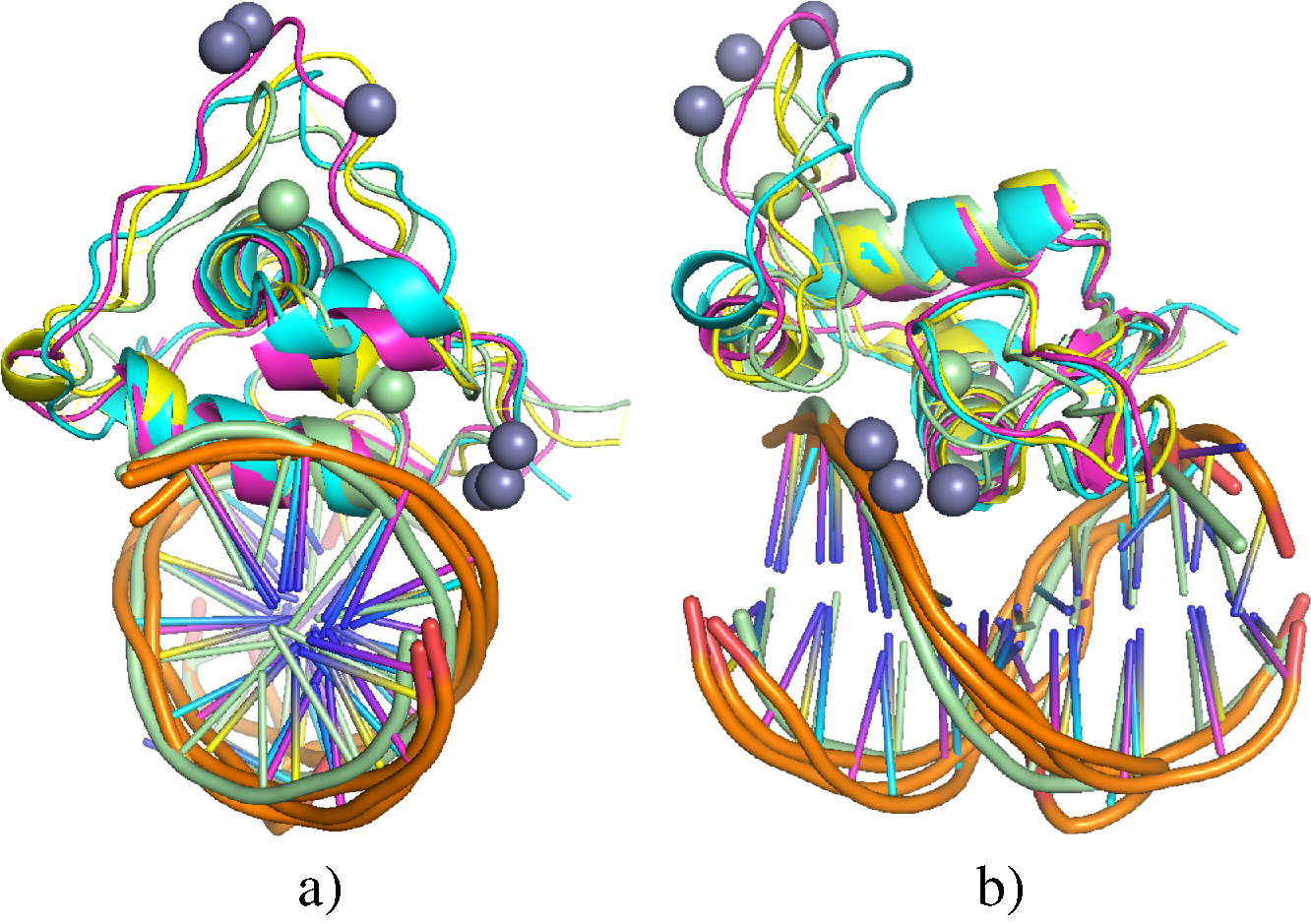}
\includegraphics[width=0.48\textwidth]{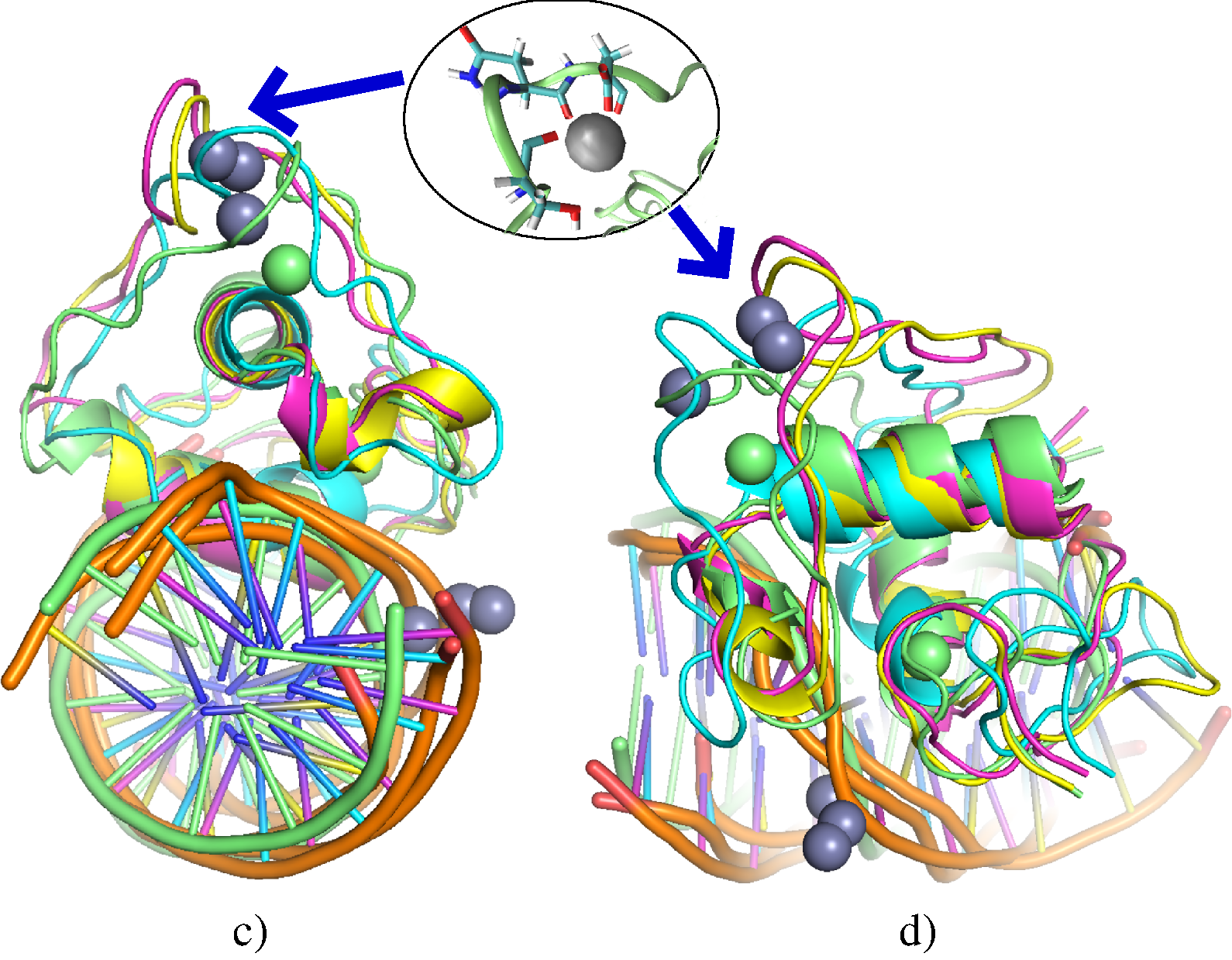}
	\caption{Comparison of the central representative configuration of the three dominant clusters 
for protomers A (a and b) and  B (c and d). 
The experimental X$-$ray structure is shown with green, and
the three configurations obtained from the simulation 
are shown in cyan, yellow, and purple.
For each protomer, the top view (along the DNA axis) and side view are presented for clarity. 
The inset in subfigures c) and d) is the``unfavorable'' dihedral angles of the
three amino acids shown in Fig. \ref{fig:rama}b that bind to zinc ion
in the new location in the first cluster.
The system is the undercharged CYS system.
	}
	\label{fig:rmsdclusterCYS}
\end{figure*}

To discern the major similarities and differences among the dominant clusters of the proteins and
to show their deviation with respect to the experimental structure,
we align and overlap the central configuration (the representative configuration) 
of these clusters. 
The results are shown in Figs. \ref{fig:rmsdclusterCYN} and \ref{fig:rmsdclusterCYS}.
It is clear that the simulated and experimental structures of the overcharged CYN system strongly overlap each other.
Nevertheless, one subtle difference is observed. While protomer A of the CYN system
keeps its structural components, protomer B of the CYN system shows appearance of additional $\beta-$ strands in 
the location where the experiment structure shows short $\beta-$bridges. 
Following the time dependent structure information shown in Fig. \ref{subfig:dsspCym}, 
one learns that these $\beta-$strands are created
after 300ns into the simulation. These $\beta-$ strands are supposed to be native to these zinc-fingers
but upon binding to DNA they are not observed in the experimental crystal structure. Our simulation
results show that the $\beta-$strands are still present, albeit transiently (blue arrow `2' in Fig. \ref{fig:dssp}a). 
This result suggests that the DNA binding of these
zinc fingers are so strong that this binding disrupts these $\beta-$strand secondary structures. In
experimental structure measurement, the temperature is effectively zero. In molecular dynamics simulation,
the temperature is finite. Hence so the $\beta-$strands have finite probability to reappear transiently.

Comparison of the simulated and experimental structures of the zinc-finger in the undercharged CYS system shown
in Fig. \ref{fig:rmsdclusterCYS}. Substantial reorganizations of the zinc ions are observed. In both protomers,
one zinc ion leaves the cysteine binding pocket and moves near the negatively charged DNA's phosphate
backbone. The other zinc ion remains with the cysteine amino acids in the loop segment of the binding
pocket, but it pushes this loop further into the water solution, far away from the DNA molecule
and leaves the helix segment of the zinc-finger behind. 
This behavior is totally understandable from the point of view of electrostatic interactions.
Since the cysteines are neutral, they only act as polarized side chains. The zinc ion binds weaker to 
neutral cysteins compared with
the CYN system. As a result, the ions have more room to explore other configurations.
The zinc ion of the zinc finger near the DNA would move to the negatively charged DNA's 
backbone to lower the electrostatic energy. The zinc ions in the zinc finger 
far away pushes toward to the water solution to enjoy a medium
with large dielectric constant and also lower its electrostatic self$-$energy. 
Despite the big movement of the zinc ions, the secondary structures
of the protomers remain relatively stable in this new configuration (albeit with larger fluctuations), 
because these structures are determined
mostly by the hydrogen bond interactions among the constituent amino acids. The most notable
change is the melting of half of the helix of protomer B in the DNA's major groove,
as shown from Fig. \ref{subfig:dsspCys}. 
Nevertheless, the helix remains in this groove throughout the simulation. 
Another notable observation is shown in the inset in Fig. \ref{fig:rmsdclusterCYS}c and 
Fig. \ref{fig:rmsdclusterCYS}d. The zinc ion in the CYS system that moved to closer 
to the aqueous solution also bound to new amino acids. 
Specifically, the ion bound to the oxygen atoms of SER580, ASN582and ASP583. These
are polar and charged amino acids, thus they also favor high dielectric constant medium, just like
the zinc ion. The electrostatic interactions between zinc and the amino acids are so strong 
and bends the dihedral angles of the peptide backbone of these amino acids into ``unfavorable'' values 
as mentioned in Fig. \ref{fig:rama}. For the protomer A, the movement of the second zinc finger
toward the solution lifts the nearby loops and helices away from the
DNA (more than protomer B). This leads to the reduction in the hydrogen bonds observed earlier.

Overall, protomers in the CYS system settle to a new equilibrium configuration with the zinc ions 
deviates substantially from their experiment positions. The protomers also show high flexibility 
meaning weaker DNA binding.
Note that the electrostatic interaction of zinc ions to the protein$-$DNA complex remains much larger
than the thermal energy due to the high valence of zinc ion ($+2$). Hence, the ions did not go into 
the solution. They are permanently displaced to the new locations in our simulation. 

\subsection{Principal component analysis and free energy landscape in collective variables}

\begin{figure*}
	\centering
	\includegraphics[width=0.98\textwidth]{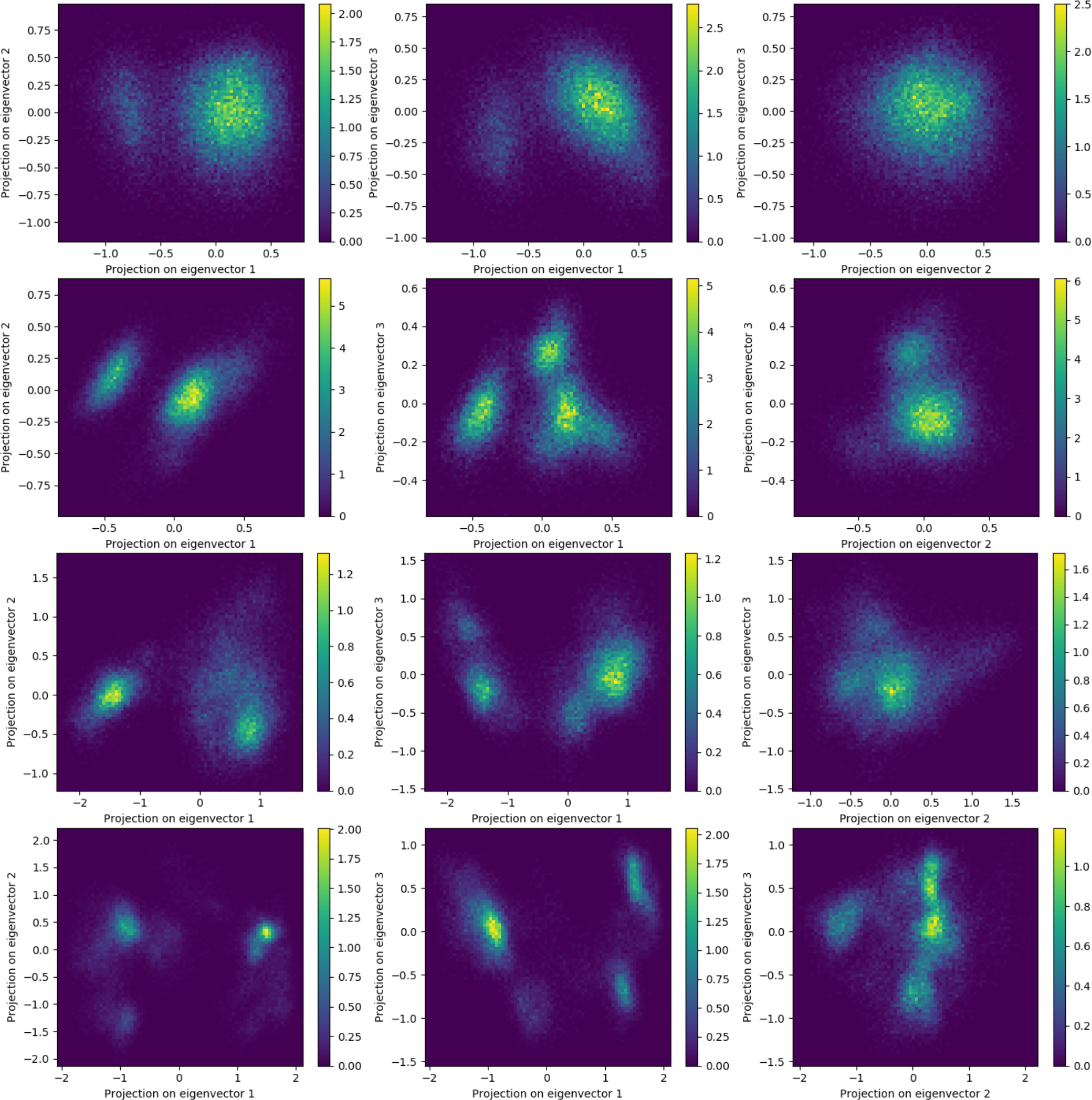}
	\caption{Normalized two-dimensional histogram of the projection of the protomers on their first 3 eigenvectors. 
Rows 1 and 2 correspond to protomers A and B in the CYN system. 
Rows 3 and 4 correspond to protomers A and B in the CYS system. 
For each row, the left figure is the projection on eigenvectors 1 and 2, the middle figure is 
the projection on eigenvectors 1 and 3, and the right figure is the projection on eigenvectors 2 and 3. 
The color bar shows the scale of the probability density amplitude and is different for different figures.}
	\label{fig:pcaall}
\end{figure*}

Principal component analysis (PCA) is a useful method to analyze the dynamical behaviors of proteins. 
Using PCA, one can screen out fast and high energy modes in the dynamics, leading
to a huge reduction in the dimensionality of the system. Just like in the case
of RMSD clustering analysis, dynamics of proteins are well described using the first
few principal collective motions of the backbone atoms. Through our own inspection, 
the three most dominant eigenvectors are enough to locate the number of distinct clusters of 
configurations of the two systems.

In Fig. \ref{fig:pcaall}, the distribution of all simulated configurations projected on the first three eigenvectors are
shown. The four rows correspond to the protomer A of the CYN system, protomer B of the CYN system, 
protomer A of the CYS system, and protomer B of the CYS system. For each row, the left, middle and right are the projections
on eigenvectors 1 and 2, 1 and 3, and 2 and 3, respectively. 
As shown on these figures, protomer A has two distinct peaks of high probabilities, 
whereas protomer B shows three peaks, once again signifying the difference
between he identical protomers upon DNA$-$binding. The influence of the charging states of the cysteine
amino acids is obvious. The peaks for the overcharged CYN system are much sharper and Gaussian-like, indicating 
structural stability, whereas the CYS system has peaks that are broader and have irregular shapes. For
CYS system, several extra small peaks appear indicating structural flexibility and
intermediate states. 

The trace of the covariance matrix of the four corresponding complexes are 0.798,
0.403, 1.467, and 1.54638 nm$^2$. The higher the value, the more structurally flexible
the system is. Therefore, one sees that protomers A and B in the CYN system are more stable
than the protomers in the CYS system, and protomer B has stronger DNA binding. 
In the CYS system, their trace values show the same flexibility indicating that weaker
DNA binding leads to less difference between protomers A and B. This behavior is expected 
because the two protomers are identical in sequence, and any difference between them 
is due to DNA binding. Thus, weaker DNA binding means less difference.

Lastly, in the coordinates of these collective variables, the free energy can be easily obtained from the probability 
density distribution function, $\Delta G \propto - k_BT \log p (\mathbf{a}_i, \mathbf{a}_j)$,
where $\mathbf{a}_i$ are projections on eigenvector $i-$th. As the color-coded values of this
Fig. \ref{fig:pcaall} shows, the CYS system has a much wider
range of these projection values leading to lower probability density distribution than the CYN system. 
Between the protomers A and B in the same system, 
protomer B shows sharper peaks and smaller range of $\mathbf{a}_i$. 
Specifically, the difference in the free energy of protomer A and B 
is $\Delta G_B - \Delta G_A = -0.94$ kJ/mol in the CYN system, whereas in the CYS system, 
the difference in the free energy between protomers A and B is less at $\Delta G_B - \Delta G_A = -0.46$ kJ/mol.
Between CYN and CYS systems, the free energy of protomer A in the CYN system 
is lower by 0.87 kJ/mol than that of protomer A in the CYS system.
Protomer B in the CYN system is lower by 1.7 kJ/mol than that in the CYS system. 
One can see from these analyses that the overcharge state is more stable with
protomer B having even lower free energy. 
In the undercharged state, the structures are more flexible. Thus,
DNA binding causes less difference in the free energy between protomers A and B. 

We can argue from these structural, dynamical and free energy analyses that strong DNA binding
is the main reason for the symmetry breaking between protomers A and B.
A full understanding of the mechanism of how DNA binding causes
the symmetry breaking between the protomers would require
much more investigation and analysis by comparing different binding poses.
For example, one can compare 
symmetric versus mirror binding poses of the two protomers to identify 
important residues and interactions forcing the symmetry breaking. 
However, the symmetric binding pose requires protein docking procedure. 
This theoretical approach that can introduce uncertainties into this delicate energy balance. 
Such studies are beyond the scope of this paper, and will be addressed in a future work.

\section{Conclusions}

In this paper, we perform a molecular dynamics simulation to investigate 
a ZnCys4 zinc finger protein dimer in its binding pose with DNA. The monomers of the dimers are
identical in sequences and bind to the same nucleic acid sequences. Yet
The two monomers have differences in structures and energies with the "downstream" complex
showing stronger binding. The overcharged state of the zinc ion is very important
for this binding. In this state, all four cysteine  amino acids are deprotonated to become
negatively charged and thereby overcharging the zinc ion. Previous works showed that
this overcharged state is important for stability of the zinc finger. In this work,
various analyses shown that this overcharged state is also very important
for the protein$-$DNA binding complex. In the undercharged state, 
the zinc ions would move to different locations in the complex to lower
their electrostatic free energy, leading to an increase in the atomic fluctuations 
and the dynamics of the complex.
Our results provide 
insights into the DNA binding state of this specific zinc finger of PSA protein
and have potential application in designing specialized biosensor 
for prostate cancer screening. 

In a broader picture of the several other zinc finger structures, one
can ask if the results of this work could be applied.
Here, we have focused on the protonation state of ZnCys4 in its binding with DNA molecules. 
Other quantum mechanical calculation studies of single ZnCys2His2 or ZnCys3His  
zinc fingers showed the cysteine amino acids are also deprotonated in binding with zinc ions 
\cite{Dudev2002,Peters2010}. 
Calculations of ZnCys2His2 zinc fingers showed that
the histidines in these structures have a
magnitude of charge transfer to zinc ions comparable with cysteines \cite{Li2008}.
Less “hacky” classical forcefield sets the histidine amino acid in zinc finger to be in a
negatively-charged deprotonated state \cite{Pang1999} for this complex to be stable. 
Our preliminary study of ZnCys3His zinc finger also shows that histidine is negatively charged for 
the complex to be stable (manuscript in preparation).
The DNA binding poses of these zinc fingers are also similar to our system. Therefore,
we believe our results 
can be applied to the elucidation of the molecular mechanism of the DNA binding of these zinc figures. 
However, beyond standard-fold $\beta\beta\alpha$ zinc finger, the binding poses of 
non-standard fold zinc fingers are different \cite{Kluska2018,Laity2001}. 
Hence, more investigations are needed for those cases
to confirm the importance of the electrostatics of zinc finger in DNA binding.

Finally, the initial experimental crystal structure of our system shows the
symmetry breaking of protomer A and B structures in binding with DNA. 
Our results show that the strong DNA binding in the overcharged state causes a strong 
difference in the free energy of protomers A and B.
In weaker uncharged state, the difference in the free energy of protomers A and B is smaller,
and their secondary structures are also more similar.
Therefore, we argue that DNA binding is the main cause of this symmetry breaking. 
However, we only focused on the electrostatics of the DNA binding of zinc finger in this work.
In practice, DNA binding properties have many aspects, and much more comprehensive studies 
of this mechanism are needed.  More elaborate simulations of different DNA binding poses
will be the subject of future works.

\begin{acknowledgement}

The authors acknowledge the financial support of the Vietnam National University, Hanoi,
grant number QG.16.01. The authors acknowledge the financial support of the World Bank
and the Ministry of Science and Technology of Vietnam grant number 
13/FIRST/1.a/VNU1. 

The authors thank Professor Morikawa Yoshitada for many useful discussions and inputs.

\end{acknowledgement}

\bibliography{zinc}

\providecommand{\latin}[1]{#1}
\makeatletter
\providecommand{\doi}
  {\begingroup\let\do\@makeother\dospecials
  \catcode`\{=1 \catcode`\}=2 \doi@aux}
\providecommand{\doi@aux}[1]{\endgroup\texttt{#1}}
\makeatother
\providecommand*\mcitethebibliography{\thebibliography}
\csname @ifundefined\endcsname{endmcitethebibliography}
  {\let\endmcitethebibliography\endthebibliography}{}
\begin{mcitethebibliography}{47}
\providecommand*\natexlab[1]{#1}
\providecommand*\mciteSetBstSublistMode[1]{}
\providecommand*\mciteSetBstMaxWidthForm[2]{}
\providecommand*\mciteBstWouldAddEndPuncttrue
  {\def\EndOfBibitem{\unskip.}}
\providecommand*\mciteBstWouldAddEndPunctfalse
  {\let\EndOfBibitem\relax}
\providecommand*\mciteSetBstMidEndSepPunct[3]{}
\providecommand*\mciteSetBstSublistLabelBeginEnd[3]{}
\providecommand*\EndOfBibitem{}
\mciteSetBstSublistMode{f}
\mciteSetBstMaxWidthForm{subitem}{(\alph{mcitesubitemcount})}
\mciteSetBstSublistLabelBeginEnd
  {\mcitemaxwidthsubitemform\space}
  {\relax}
  {\relax}

\bibitem[Lander \latin{et~al.}(2001)Lander, Linton, Birren, Nusbaum, Zody,
  Baldwin, Devon, Dewar, Doyle, FitzHugh, Funke, Gage, Harris, Heaford,
  Howland, Kann, Lehoczky, LeVine, McEwan, McKernan, Meldrim, Mesirov, Miranda,
  Morris, Naylor, Raymond, Rosetti, Santos, Sheridan, Sougnez, Stange-Thomann,
  Stojanovic, Subramanian, Wyman, Rogers, Sulston, Ainscough, Beck, Bentley,
  Burton, Clee, Carter, Coulson, Deadman, Deloukas, Dunham, Dunham, Durbin,
  French, Grafham, Gregory, Hubbard, Humphray, Hunt, Jones, Lloyd, McMurray,
  Matthews, Mercer, Milne, Mullikin, Mungall, Plumb, Ross, Shownkeen, Sims,
  Waterston, Wilson, Hillier, McPherson, Marra, Mardis, Fulton, Chinwalla,
  Pepin, Gish, Chissoe, Wendl, Delehaunty, Miner, Delehaunty, Kramer, Cook,
  Fulton, Johnson, Minx, Clifton, Hawkins, Branscomb, Predki, Richardson,
  Wenning, Slezak, Doggett, Cheng, Olsen, Lucas, Elkin, Uberbacher, Frazier,
  Gibbs, Muzny, Scherer, Bouck, Sodergren, Worley, Rives, Gorrell, Metzker,
  Naylor, Kucherlapati, Nelson, Weinstock, Sakaki, Fujiyama, Hattori, Yada,
  Toyoda, Itoh, Kawagoe, Watanabe, Totoki, Taylor, Weissenbach, Heilig, Saurin,
  Artiguenave, Brottier, Bruls, Pelletier, Robert, Wincker, Rosenthal, Platzer,
  Nyakatura, Taudien, Rump, Smith, Doucette-Stamm, Rubenfield, Weinstock, Lee,
  Dubois, Yang, Yu, Wang, Huang, Gu, Hood, Rowen, Madan, Qin, Davis,
  Federspiel, Abola, Proctor, Roe, Chen, Pan, Ramser, Lehrach, Reinhardt,
  McCombie, de~la Bastide, Dedhia, Blöcker, Hornischer, Nordsiek, Agarwala,
  Aravind, Bailey, Bateman, Batzoglou, Birney, Bork, Brown, Burge, Cerutti,
  Chen, Church, Clamp, Copley, Doerks, Eddy, Eichler, Furey, Galagan, Gilbert,
  Harmon, Hayashizaki, Haussler, Hermjakob, Hokamp, Jang, Johnson, Jones,
  Kasif, Kaspryzk, Kennedy, Kent, Kitts, Koonin, Korf, Kulp, Lancet, Lowe,
  McLysaght, Mikkelsen, Moran, Mulder, Pollara, Ponting, Schuler, Schultz,
  Slater, Smit, Stupka, Szustakowki, Thierry-Mieg, Thierry-Mieg, Wagner,
  Wallis, Wheeler, Williams, Wolf, Wolfe, Yang, Yeh, Collins, Guyer, Peterson,
  Felsenfeld, Wetterstrand, Myers, Schmutz, Dickson, Grimwood, Cox, Olson,
  Kaul, Raymond, Shimizu, Kawasaki, Minoshima, Evans, Athanasiou, Schultz,
  Patrinos, Morgan, Consortium, Whitehead Institute~for Biomedical~Research,
  Centre:, Center, Institute:, of~Medicine Human Genome Sequencing~Center:,
  Center:, Genoscope, UMR-8030:, Department~of Genome~Analysis, Center:,
  Center:, Multimegabase Sequencing~Center, Center:, of~Oklahoma's Advanced
  Center~for Genome~Technology:, for Molecular~Genetics:, Cold Spring
  Harbor~Laboratory, for Biotechnology:, *Genome Analysis Group (listed
  in~alphabetical order, Scientific~management: National Human Genome
  Research~Institute, Center:, of~Washington Genome~Center:, Department~of
  Molecular~Biology, of~Texas Southwestern Medical Center~at Dallas:, Office~of
  Science, and Trust:]{Lander2001}
Lander,~E.~S. \latin{et~al.}  (2001) Initial sequencing and analysis of the
  human genome. \emph{Nature} \emph{409}, 860--921\relax
\mciteBstWouldAddEndPuncttrue
\mciteSetBstMidEndSepPunct{\mcitedefaultmidpunct}
{\mcitedefaultendpunct}{\mcitedefaultseppunct}\relax
\EndOfBibitem
\bibitem[Klug(2010)]{Klug2010}
Klug,~A. (2010) The Discovery of Zinc Fingers and Their Applications in Gene
  Regulation and Genome Manipulation. \emph{Annual Review of Biochemistry}
  \emph{79}, 213--231\relax
\mciteBstWouldAddEndPuncttrue
\mciteSetBstMidEndSepPunct{\mcitedefaultmidpunct}
{\mcitedefaultendpunct}{\mcitedefaultseppunct}\relax
\EndOfBibitem
\bibitem[Kluska \latin{et~al.}(2018)Kluska, Adamczyk, and
  Kr{\k{e}}{\.{z}}el]{Kluska2018}
Kluska,~K., Adamczyk,~J., and Kr{\k{e}}{\.{z}}el,~A. (2018) Metal binding
  properties, stability and reactivity of zinc fingers. \emph{Coordination
  Chemistry Reviews} \emph{367}, 18--64\relax
\mciteBstWouldAddEndPuncttrue
\mciteSetBstMidEndSepPunct{\mcitedefaultmidpunct}
{\mcitedefaultendpunct}{\mcitedefaultseppunct}\relax
\EndOfBibitem
\bibitem[Jen and Wang(2016)Jen, and Wang]{jen2016}
Jen,~J., and Wang,~Y.-C. (2016) Zinc finger proteins in cancer progression.
  \emph{Journal of Biomedical Science} \emph{23}, 53\relax
\mciteBstWouldAddEndPuncttrue
\mciteSetBstMidEndSepPunct{\mcitedefaultmidpunct}
{\mcitedefaultendpunct}{\mcitedefaultseppunct}\relax
\EndOfBibitem
\bibitem[Mitra \latin{et~al.}(2013)Mitra, Wang, Vo, Rouzina, Barany, and
  Musier-Forsyth]{Mitra2013}
Mitra,~M., Wang,~W., Vo,~M.-N., Rouzina,~I., Barany,~G., and Musier-Forsyth,~K.
  (2013) The N-Terminal Zinc Finger and Flanking Basic Domains Represent the
  Minimal Region of the Human Immunodeficiency Virus Type-1 Nucleocapsid
  Protein for Targeting Chaperone Function. \emph{Biochemistry} \emph{52},
  8226--8236\relax
\mciteBstWouldAddEndPuncttrue
\mciteSetBstMidEndSepPunct{\mcitedefaultmidpunct}
{\mcitedefaultendpunct}{\mcitedefaultseppunct}\relax
\EndOfBibitem
\bibitem[Jamieson \latin{et~al.}(2003)Jamieson, Miller, and Pabo]{Jamieson2003}
Jamieson,~A.~C., Miller,~J.~C., and Pabo,~C.~O. (2003) Drug discovery with
  engineered zinc-finger proteins. \emph{Nature Reviews Drug Discovery}
  \emph{2}, 361--368\relax
\mciteBstWouldAddEndPuncttrue
\mciteSetBstMidEndSepPunct{\mcitedefaultmidpunct}
{\mcitedefaultendpunct}{\mcitedefaultseppunct}\relax
\EndOfBibitem
\bibitem[Balk \latin{et~al.}(2003)Balk, Ko, and Bubley]{balk2003biology}
Balk,~S.~P., Ko,~Y.-J., and Bubley,~G.~J. (2003) Biology of prostate-specific
  antigen. \emph{J. Clin. Oncol.} \emph{21}, 383--391\relax
\mciteBstWouldAddEndPuncttrue
\mciteSetBstMidEndSepPunct{\mcitedefaultmidpunct}
{\mcitedefaultendpunct}{\mcitedefaultseppunct}\relax
\EndOfBibitem
\bibitem[Formisano \latin{et~al.}(2015)Formisano, Jolly, Bhalla, Cromhout,
  Flanagan, Fogel, Limson, and Estrela]{formisano2015optimisation}
Formisano,~N., Jolly,~P., Bhalla,~N., Cromhout,~M., Flanagan,~S.~P., Fogel,~R.,
  Limson,~J.~L., and Estrela,~P. (2015) Optimisation of an electrochemical
  impedance spectroscopy aptasensor by exploiting quartz crystal microbalance
  with dissipation signals. \emph{Sens. Actuators B Chem.} \emph{220},
  369--375\relax
\mciteBstWouldAddEndPuncttrue
\mciteSetBstMidEndSepPunct{\mcitedefaultmidpunct}
{\mcitedefaultendpunct}{\mcitedefaultseppunct}\relax
\EndOfBibitem
\bibitem[Aus \latin{et~al.}(2005)Aus, Damber, Khatami, Lilja, Stranne, and
  Hugosson]{aus2005individualized}
Aus,~G., Damber,~J.-E., Khatami,~A., Lilja,~H., Stranne,~J., and Hugosson,~J.
  (2005) Individualized screening interval for prostate cancer based on
  prostate-specific antigen level: results of a prospective, randomized,
  population-based study. \emph{Arch. Intern. Med.} \emph{165},
  1857--1861\relax
\mciteBstWouldAddEndPuncttrue
\mciteSetBstMidEndSepPunct{\mcitedefaultmidpunct}
{\mcitedefaultendpunct}{\mcitedefaultseppunct}\relax
\EndOfBibitem
\bibitem[Botchorishvili \latin{et~al.}(2009)Botchorishvili, Matikainen, and
  Lilja]{botchorishvili2009}
Botchorishvili,~G., Matikainen,~M.~P., and Lilja,~H. (2009) Early
  prostate-specific antigen changes and the diagnosis and prognosis of prostate
  cancer. \emph{Current Opinion in Urology} \emph{19}, 221--226\relax
\mciteBstWouldAddEndPuncttrue
\mciteSetBstMidEndSepPunct{\mcitedefaultmidpunct}
{\mcitedefaultendpunct}{\mcitedefaultseppunct}\relax
\EndOfBibitem
\bibitem[Lilja \latin{et~al.}(2008)Lilja, Ulmert, and Vickers]{lilja2008}
Lilja,~H., Ulmert,~D., and Vickers,~A.~J. (2008) Prostate-specific antigen and
  prostate cancer: prediction, detection and monitoring. \emph{Nature Reviews
  Cancer} \emph{8}, 268--278\relax
\mciteBstWouldAddEndPuncttrue
\mciteSetBstMidEndSepPunct{\mcitedefaultmidpunct}
{\mcitedefaultendpunct}{\mcitedefaultseppunct}\relax
\EndOfBibitem
\bibitem[Liu \latin{et~al.}(2012)Liu, Lu, Hua, Jiang, and
  Xie]{liu2012detection}
Liu,~B., Lu,~L., Hua,~E., Jiang,~S., and Xie,~G. (2012) Detection of the human
  prostate-specific antigen using an aptasensor with gold nanoparticles
  encapsulated by graphitized mesoporous carbon. \emph{Microchim. Acta}
  \emph{178}, 163--170\relax
\mciteBstWouldAddEndPuncttrue
\mciteSetBstMidEndSepPunct{\mcitedefaultmidpunct}
{\mcitedefaultendpunct}{\mcitedefaultseppunct}\relax
\EndOfBibitem
\bibitem[Lai \latin{et~al.}(2006)Lai, Kedda, Hinze, Smith, Yaxley, Spurdle,
  Morris, Harris, and Clements]{lai2006}
Lai,~J., Kedda,~M.-A., Hinze,~K., Smith,~R.~L., Yaxley,~J., Spurdle,~A.~B.,
  Morris,~C., Harris,~J., and Clements,~J.~A. (2006) {PSA}/{KLK}3 {AREI}
  promoter polymorphism alters androgen receptor binding and is associated with
  prostate cancer susceptibility. \emph{Carcinogenesis} \emph{28},
  1032--1039\relax
\mciteBstWouldAddEndPuncttrue
\mciteSetBstMidEndSepPunct{\mcitedefaultmidpunct}
{\mcitedefaultendpunct}{\mcitedefaultseppunct}\relax
\EndOfBibitem
\bibitem[Laity \latin{et~al.}(2001)Laity, Lee, and Wright]{Laity2001}
Laity,~J.~H., Lee,~B.~M., and Wright,~P.~E. (2001) Zinc finger proteins: new
  insights into structural and functional diversity. \emph{Current Opinion in
  Structural Biology} \emph{11}, 39--46\relax
\mciteBstWouldAddEndPuncttrue
\mciteSetBstMidEndSepPunct{\mcitedefaultmidpunct}
{\mcitedefaultendpunct}{\mcitedefaultseppunct}\relax
\EndOfBibitem
\bibitem[Wolfe \latin{et~al.}(2000)Wolfe, Nekludova, and Pabo]{Wolfe2000}
Wolfe,~S.~A., Nekludova,~L., and Pabo,~C.~O. (2000) {DNA} Recognition by
  Cys2His2Zinc Finger Proteins. \emph{Annual Review of Biophysics and
  Biomolecular Structure} \emph{29}, 183--212\relax
\mciteBstWouldAddEndPuncttrue
\mciteSetBstMidEndSepPunct{\mcitedefaultmidpunct}
{\mcitedefaultendpunct}{\mcitedefaultseppunct}\relax
\EndOfBibitem
\bibitem[Shaffer \latin{et~al.}(2004)Shaffer, Jivan, Dollins, Claessens, and
  Gewirth]{shaffer2004}
Shaffer,~P.~L., Jivan,~A., Dollins,~D.~E., Claessens,~F., and Gewirth,~D.~T.
  (2004) Structural basis of androgen receptor binding to selective androgen
  response elements. \emph{Proceedings of the National Academy of Sciences}
  \emph{101}, 4758--4763\relax
\mciteBstWouldAddEndPuncttrue
\mciteSetBstMidEndSepPunct{\mcitedefaultmidpunct}
{\mcitedefaultendpunct}{\mcitedefaultseppunct}\relax
\EndOfBibitem
\bibitem[Godwin \latin{et~al.}(2017)Godwin, Melvin, Gmeiner, and
  Salsbury]{Godwin_2017}
Godwin,~R.~C., Melvin,~R.~L., Gmeiner,~W.~H., and Salsbury,~F.~R. (2017)
  Binding Site Configurations Probe the Structure and Dynamics of the Zinc
  Finger of {NEMO} ({NF}-$\upkappa$B Essential Modulator). \emph{Biochemistry}
  \emph{56}, 623--633\relax
\mciteBstWouldAddEndPuncttrue
\mciteSetBstMidEndSepPunct{\mcitedefaultmidpunct}
{\mcitedefaultendpunct}{\mcitedefaultseppunct}\relax
\EndOfBibitem
\bibitem[Godwin \latin{et~al.}(2015)Godwin, Gmeiner, and Salsbury]{Godwin2015}
Godwin,~R., Gmeiner,~W., and Salsbury,~F.~R. (2015) Importance of long-time
  simulations for rare event sampling in zinc finger proteins. \emph{Journal of
  Biomolecular Structure and Dynamics} \emph{34}, 125--134\relax
\mciteBstWouldAddEndPuncttrue
\mciteSetBstMidEndSepPunct{\mcitedefaultmidpunct}
{\mcitedefaultendpunct}{\mcitedefaultseppunct}\relax
\EndOfBibitem
\bibitem[Hamed and Arya(2016)Hamed, and Arya]{Hamed2016}
Hamed,~M.~Y., and Arya,~G. (2016) Zinc finger protein binding to {DNA}: an
  energy perspective using molecular dynamics simulation and free energy
  calculations on mutants of both zinc finger domains and their specific {DNA}
  bases. \emph{Journal of Biomolecular Structure and Dynamics} \emph{34},
  919--934\relax
\mciteBstWouldAddEndPuncttrue
\mciteSetBstMidEndSepPunct{\mcitedefaultmidpunct}
{\mcitedefaultendpunct}{\mcitedefaultseppunct}\relax
\EndOfBibitem
\bibitem[Lee \latin{et~al.}(2010)Lee, Kim, and Seok]{Lee2010a}
Lee,~J., Kim,~J.-S., and Seok,~C. (2010) Cooperativity and Specificity of
  Cys2His2Zinc Finger Protein-{DNA} Interactions: A Molecular Dynamics
  Simulation Study. \emph{The Journal of Physical Chemistry B} \emph{114},
  7662--7671\relax
\mciteBstWouldAddEndPuncttrue
\mciteSetBstMidEndSepPunct{\mcitedefaultmidpunct}
{\mcitedefaultendpunct}{\mcitedefaultseppunct}\relax
\EndOfBibitem
\bibitem[Alberts \latin{et~al.}(2014)Alberts, Johnson, Lewis, Morgan, Raff,
  Roberts, and Walter]{cellBook}
Alberts,~B., Johnson,~A.~D., Lewis,~J., Morgan,~D., Raff,~M., Roberts,~K., and
  Walter,~P. \emph{Molecular Biology of the Cell}, 6th ed.; W. W. Norton \&
  Company, 2014\relax
\mciteBstWouldAddEndPuncttrue
\mciteSetBstMidEndSepPunct{\mcitedefaultmidpunct}
{\mcitedefaultendpunct}{\mcitedefaultseppunct}\relax
\EndOfBibitem
\bibitem[Lee \latin{et~al.}(2011)Lee, Tran, and Nguyen]{Lee2011}
Lee,~S., Tran,~C.~V., and Nguyen,~T.~T. (2011) Inhibition of DNA ejection from
  bacteriophage by Mg$^{+2}$ counterions. \emph{J. Chem Phys.} \emph{134},
  125104\relax
\mciteBstWouldAddEndPuncttrue
\mciteSetBstMidEndSepPunct{\mcitedefaultmidpunct}
{\mcitedefaultendpunct}{\mcitedefaultseppunct}\relax
\EndOfBibitem
\bibitem[Dudev and Lim(2002)Dudev, and Lim]{Dudev2002}
Dudev,~T., and Lim,~C. (2002) Factors Governing the Protonation State of
  Cysteines in Proteins:~ An Ab Initio/{CDM} Study. \emph{Journal of the
  American Chemical Society} \emph{124}, 6759--6766\relax
\mciteBstWouldAddEndPuncttrue
\mciteSetBstMidEndSepPunct{\mcitedefaultmidpunct}
{\mcitedefaultendpunct}{\mcitedefaultseppunct}\relax
\EndOfBibitem
\bibitem[Peters \latin{et~al.}(2010)Peters, Yang, Wang, F\"{u}sti-Moln{\'{a}}r,
  Weaver, and Merz]{Peters2010}
Peters,~M.~B., Yang,~Y., Wang,~B., F\"{u}sti-Moln{\'{a}}r,~L., Weaver,~M.~N.,
  and Merz,~K.~M. (2010) Structural Survey of Zinc-Containing Proteins and
  Development of the Zinc {AMBER} Force Field ({ZAFF}). \emph{Journal of
  Chemical Theory and Computation} \emph{6}, 2935--2947\relax
\mciteBstWouldAddEndPuncttrue
\mciteSetBstMidEndSepPunct{\mcitedefaultmidpunct}
{\mcitedefaultendpunct}{\mcitedefaultseppunct}\relax
\EndOfBibitem
\bibitem[Brandt \latin{et~al.}(2009)Brandt, Hellgren, Brinck, Bergman, and
  Edholm]{Brandt2009}
Brandt,~E.~G., Hellgren,~M., Brinck,~T., Bergman,~T., and Edholm,~O. (2009)
  Molecular dynamics study of zinc binding to cysteines in a peptide mimic of
  the alcohol dehydrogenase structural zinc site. \emph{Phys. Chem. Chem.
  Phys.} \emph{11}, 975--983\relax
\mciteBstWouldAddEndPuncttrue
\mciteSetBstMidEndSepPunct{\mcitedefaultmidpunct}
{\mcitedefaultendpunct}{\mcitedefaultseppunct}\relax
\EndOfBibitem
\bibitem[Grosberg \latin{et~al.}(2002)Grosberg, Nguyen, and
  Shklovskii]{Grosberg2002}
Grosberg,~A.~Y., Nguyen,~T.~T., and Shklovskii,~B. (2002) Low temperature
  physics at room temperature in water: Charge inversion in chemical and
  biological systems. \emph{Rev. Mod. Phys.} \emph{74}, 329--345\relax
\mciteBstWouldAddEndPuncttrue
\mciteSetBstMidEndSepPunct{\mcitedefaultmidpunct}
{\mcitedefaultendpunct}{\mcitedefaultseppunct}\relax
\EndOfBibitem
\bibitem[Nguyen \latin{et~al.}(2017)Nguyen, Nguyen, and Carloni]{Nguyen2017}
Nguyen,~V.~D., Nguyen,~T.~T., and Carloni,~P. (2017) {DNA} like-charge
  attraction and overcharging by divalent counterions in the presence of
  divalent co-ions. \emph{Journal of Biological Physics} \emph{43},
  185--195\relax
\mciteBstWouldAddEndPuncttrue
\mciteSetBstMidEndSepPunct{\mcitedefaultmidpunct}
{\mcitedefaultendpunct}{\mcitedefaultseppunct}\relax
\EndOfBibitem
\bibitem[Nguyen(2016)]{Nguyen2016}
Nguyen,~T.~T. (2016) Grand-canonical simulation of {DNA} condensation with two
  salts, effect of divalent counterion size. \emph{The Journal of Chemical
  Physics} \emph{144}, 065102\relax
\mciteBstWouldAddEndPuncttrue
\mciteSetBstMidEndSepPunct{\mcitedefaultmidpunct}
{\mcitedefaultendpunct}{\mcitedefaultseppunct}\relax
\EndOfBibitem
\bibitem[Hall \latin{et~al.}(2009)Hall, van Dorp, Lemay, and Dekker]{Hall2009}
Hall,~A.~R., van Dorp,~S., Lemay,~S.~G., and Dekker,~C. (2009) Electrophoretic
  Force on a Protein-Coated {DNA} Molecule in a Solid-State Nanopore.
  \emph{Nano Letters} \emph{9}, 4441--4445\relax
\mciteBstWouldAddEndPuncttrue
\mciteSetBstMidEndSepPunct{\mcitedefaultmidpunct}
{\mcitedefaultendpunct}{\mcitedefaultseppunct}\relax
\EndOfBibitem
\bibitem[Netz(2001)]{Netz2001}
Netz,~R.~R. (2001) Electrostatistics of counter$-$ions at and between planar
  charged walls: From Poisson$-$Boltzmann to the strong-coupling theory.
  \emph{Eur. Phys. J. E} \emph{5}, 557--574\relax
\mciteBstWouldAddEndPuncttrue
\mciteSetBstMidEndSepPunct{\mcitedefaultmidpunct}
{\mcitedefaultendpunct}{\mcitedefaultseppunct}\relax
\EndOfBibitem
\bibitem[Gelbart \latin{et~al.}(2000)Gelbart, Bruinsma, Pincus, and
  Parsegian]{Gelbart2000}
Gelbart,~W.~M., Bruinsma,~R.~F., Pincus,~P.~A., and Parsegian,~V.~A. (2000)
  {DNA}-Inspired Electrostatics. \emph{Physics Today} \emph{53}, 38--44\relax
\mciteBstWouldAddEndPuncttrue
\mciteSetBstMidEndSepPunct{\mcitedefaultmidpunct}
{\mcitedefaultendpunct}{\mcitedefaultseppunct}\relax
\EndOfBibitem
\bibitem[Gr{\o}nbech-Jensen \latin{et~al.}(1997)Gr{\o}nbech-Jensen, Mashl,
  Bruinsma, and Gelbart]{GronbechJensen1997a}
Gr{\o}nbech-Jensen,~N., Mashl,~R.~J., Bruinsma,~R.~F., and Gelbart,~W.~M.
  (1997) Counterion-Induced Attraction between Rigid Polyelectrolytes.
  \emph{Physical Review Letters} \emph{78}, 2477--2480\relax
\mciteBstWouldAddEndPuncttrue
\mciteSetBstMidEndSepPunct{\mcitedefaultmidpunct}
{\mcitedefaultendpunct}{\mcitedefaultseppunct}\relax
\EndOfBibitem
\bibitem[Naji \latin{et~al.}(2004)Naji, Arnold, Holm, and Netz]{Naji2004}
Naji,~A., Arnold,~A., Holm,~C., and Netz,~R.~R. (2004) Attraction and unbinding
  of like-charged rods. \emph{Eur. Phys. Lett.} \emph{67}, 130--136\relax
\mciteBstWouldAddEndPuncttrue
\mciteSetBstMidEndSepPunct{\mcitedefaultmidpunct}
{\mcitedefaultendpunct}{\mcitedefaultseppunct}\relax
\EndOfBibitem
\bibitem[Showalter and Br{\"u}schweiler(2007)Showalter, and
  Br{\"u}schweiler]{showalter2007validation}
Showalter,~S.~A., and Br{\"u}schweiler,~R. (2007) Validation of molecular
  dynamics simulations of biomolecules using NMR spin relaxation as benchmarks:
  application to the AMBER99SB force field. \emph{J. Chem. Theory Comput.}
  \emph{3}, 961--975\relax
\mciteBstWouldAddEndPuncttrue
\mciteSetBstMidEndSepPunct{\mcitedefaultmidpunct}
{\mcitedefaultendpunct}{\mcitedefaultseppunct}\relax
\EndOfBibitem
\bibitem[Ivani \latin{et~al.}(2016)Ivani, Dans, Noy, Pérez, Faustino,
  Hospital, Walther, Andrio, Goñi, Balaceanu, Portella, Battistini, Gelpí,
  González, Vendruscolo, Laughton, Harris, Case, and Orozco]{Ivani2016}
Ivani,~I. \latin{et~al.}  (2016) Parmbsc1: a refined force field for DNA
  simulations. \emph{Nature Methods} \emph{13}, 55--58\relax
\mciteBstWouldAddEndPuncttrue
\mciteSetBstMidEndSepPunct{\mcitedefaultmidpunct}
{\mcitedefaultendpunct}{\mcitedefaultseppunct}\relax
\EndOfBibitem
\bibitem[Price and Brooks~III(2004)Price, and Brooks~III]{price2004modified}
Price,~D.~J., and Brooks~III,~C.~L. (2004) A modified TIP3P water potential for
  simulation with Ewald summation. \emph{J. Chem. Phys.} \emph{121},
  10096--10103\relax
\mciteBstWouldAddEndPuncttrue
\mciteSetBstMidEndSepPunct{\mcitedefaultmidpunct}
{\mcitedefaultendpunct}{\mcitedefaultseppunct}\relax
\EndOfBibitem
\bibitem[Hess \latin{et~al.}(2008)Hess, Kutzner, Van Der~Spoel, and
  Lindahl]{hess2008gromacs}
Hess,~B., Kutzner,~C., Van Der~Spoel,~D., and Lindahl,~E. (2008) GROMACS 4:
  algorithms for highly efficient, load-balanced, and scalable molecular
  simulation. \emph{J. Chem. Theory Comput.} \emph{4}, 435--447\relax
\mciteBstWouldAddEndPuncttrue
\mciteSetBstMidEndSepPunct{\mcitedefaultmidpunct}
{\mcitedefaultendpunct}{\mcitedefaultseppunct}\relax
\EndOfBibitem
\bibitem[Nos{\'e}(1984)]{nose1984molecular}
Nos{\'e},~S. (1984) A molecular dynamics method for simulations in the
  canonical ensemble. \emph{Mol. Phys.} \emph{52}, 255--268\relax
\mciteBstWouldAddEndPuncttrue
\mciteSetBstMidEndSepPunct{\mcitedefaultmidpunct}
{\mcitedefaultendpunct}{\mcitedefaultseppunct}\relax
\EndOfBibitem
\bibitem[Hoover(1985)]{hoover1985canonical}
Hoover,~W.~G. (1985) Canonical dynamics: Equilibrium phase-space distributions.
  \emph{Phys. Rev. A} \emph{31}, 1695--1697\relax
\mciteBstWouldAddEndPuncttrue
\mciteSetBstMidEndSepPunct{\mcitedefaultmidpunct}
{\mcitedefaultendpunct}{\mcitedefaultseppunct}\relax
\EndOfBibitem
\bibitem[Parrinello and Rahman(1980)Parrinello, and Rahman]{parrinello1980m}
Parrinello,~M., and Rahman,~A. (1980) Crystal Structure and Pair Potentials: A
  Molecular-Dynamics Study. \emph{Physical Review Letters} \emph{45},
  1196--1199\relax
\mciteBstWouldAddEndPuncttrue
\mciteSetBstMidEndSepPunct{\mcitedefaultmidpunct}
{\mcitedefaultendpunct}{\mcitedefaultseppunct}\relax
\EndOfBibitem
\bibitem[Parrinello and Rahman(1981)Parrinello, and
  Rahman]{parrinello1981polymorphic}
Parrinello,~M., and Rahman,~A. (1981) Polymorphic transitions in single
  crystals: A new molecular dynamics method. \emph{J. Appl. Phys.} \emph{52},
  7182--7190\relax
\mciteBstWouldAddEndPuncttrue
\mciteSetBstMidEndSepPunct{\mcitedefaultmidpunct}
{\mcitedefaultendpunct}{\mcitedefaultseppunct}\relax
\EndOfBibitem
\bibitem[Darden \latin{et~al.}(1993)Darden, York, and
  Pedersen]{darden1993particle}
Darden,~T., York,~D., and Pedersen,~L. (1993) Particle mesh Ewald: An N log (N)
  method for Ewald sums in large systems. \emph{J. Chem. Phys.} \emph{98},
  10089--10092\relax
\mciteBstWouldAddEndPuncttrue
\mciteSetBstMidEndSepPunct{\mcitedefaultmidpunct}
{\mcitedefaultendpunct}{\mcitedefaultseppunct}\relax
\EndOfBibitem
\bibitem[Hess \latin{et~al.}(1997)Hess, Bekker, Berendsen, and
  Fraaije]{hess1997lincs}
Hess,~B., Bekker,~H., Berendsen,~H.~J., and Fraaije,~J.~G. (1997) LINCS: a
  linear constraint solver for molecular simulations. \emph{J. Comput. Chem.}
  \emph{18}, 1463--1472\relax
\mciteBstWouldAddEndPuncttrue
\mciteSetBstMidEndSepPunct{\mcitedefaultmidpunct}
{\mcitedefaultendpunct}{\mcitedefaultseppunct}\relax
\EndOfBibitem
\bibitem[Humphrey \latin{et~al.}(1996)Humphrey, Dalke, and
  Schulten]{humphrey1996vmd}
Humphrey,~W., Dalke,~A., and Schulten,~K. (1996) VMD: visual molecular
  dynamics. \emph{J Mol Graph} \emph{14}, 33--38\relax
\mciteBstWouldAddEndPuncttrue
\mciteSetBstMidEndSepPunct{\mcitedefaultmidpunct}
{\mcitedefaultendpunct}{\mcitedefaultseppunct}\relax
\EndOfBibitem
\bibitem[Li \latin{et~al.}(2008)Li, Zhang, Wang, and Wang]{Li2008}
Li,~W., Zhang,~J., Wang,~J., and Wang,~W. (2008) Metal-Coupled Folding of
  Cys2His2 Zinc-Finger. \emph{Journal of the American Chemical Society}
  \emph{130}, 892--900\relax
\mciteBstWouldAddEndPuncttrue
\mciteSetBstMidEndSepPunct{\mcitedefaultmidpunct}
{\mcitedefaultendpunct}{\mcitedefaultseppunct}\relax
\EndOfBibitem
\bibitem[Pang(1999)]{Pang1999}
Pang,~Y.-P. (1999) Novel Zinc Protein Molecular Dynamics Simulations: Steps
  Toward Antiangiogenesis for Cancer Treatment. \emph{Journal of Molecular
  Modeling} \emph{5}, 196--202\relax
\mciteBstWouldAddEndPuncttrue
\mciteSetBstMidEndSepPunct{\mcitedefaultmidpunct}
{\mcitedefaultendpunct}{\mcitedefaultseppunct}\relax
\EndOfBibitem
\end{mcitethebibliography}

\end{document}